\documentclass[12pt]{iopart}
\usepackage{graphicx}
\usepackage{iopams}
\usepackage{multirow}
\usepackage{enumitem}
\usepackage{ulem}
\begin{document}

\title[MLWFs in LaMnO$_{3}$ using PBE, PBE$+U$, HSE and GW$_0$]
  {Maximally localized Wannier functions in LaMnO$_3$ within PBE$+U$,
  hybrid functionals, and partially self-consistent GW: an efficient route to construct ab-initio
  tight-binding parameters for $e_g$ perovskites}

\author{C. Franchini}
\address{Faculty of Physics, University of Vienna and Center for Computational Materials Science, A-1090 Wien, Austria.}
\ead{cesare.franchini@univie.ac.at}

\author{R. Kov\'a\v{c}ik}
\address{School of Physics, Trinity College Dublin, Dublin 2, Ireland.}
\address{Peter Gr\"unberg Institut and Institute for Advanced Simulation, Forschungszentrum J\"ulich, D-52425 J\"ulich, Germany.}
\ead{r.kovacik@fz-juelich.de}

\author{M. Marsman}
\address{Faculty of Physics, University of Vienna and Center for Computational Materials Science, A-1090 Wien, Austria.}

\author{S. Sathyanarayana Murthy}
\address{Faculty of Physics, University of Vienna and Center for Computational Materials Science, A-1090 Wien, Austria.}

\author{J. He}
\address{Faculty of Physics, University of Vienna and Center for Computational Materials Science, A-1090 Wien, Austria.}

\author{C. Ederer}
\address{School of Physics, Trinity College Dublin, Dublin 2, Ireland.}
\address{Materials Theory, ETH Z\"urich, Wolfgang-Pauli-Strasse 27, 8093 Z\"urich, Switzerland.}

\author{G. Kresse}
\address{Faculty of Physics, University of Vienna and Center for Computational Materials Science, A-1090 Wien, Austria.}


\begin{abstract}
  Using the newly developed VASP2WANNIER90 interface we have
  constructed maximally localized Wannier functions (MLWFs) for the
  $e_g$ states of the prototypical Jahn-Teller magnetic perovskite
  LaMnO$_3$ at different levels of approximation for the
  exchange-correlation kernel. These include conventional density
  functional theory (DFT) with and without additional on-site Hubbard
  $U$ term, hybrid-DFT, and partially self-consistent GW. By suitably mapping the
  MLWFs onto an effective $e_g$ tight-binding (TB) Hamiltonian we have
  computed a complete set of TB parameters which should serve as
  guidance for more elaborate treatments of correlation effects in
  effective Hamiltonian-based approaches. The method-dependent changes
  of the calculated TB parameters and their interplay with the
  electron-electron (el-el) interaction term are discussed and
  interpreted. We discuss two alternative model parameterizations: one
  in which the effects of the el-el interaction are implicitly
  incorporated in the otherwise ``noninteracting'' TB parameters, and
  a second where we include an explicit mean-field el-el interaction
  term in the TB Hamiltonian. Both models yield a set of tabulated TB
  parameters which provide the band dispersion in excellent agreement
  with the underlying {\em ab initio} and MLWF bands.
\end{abstract}

\maketitle

------------------------------------------------------------------------------------------------     
\section{\label{sec:intro}Introduction}
------------------------------------------------------------------------------------------------     

\begin{figure}
  \centering
  \includegraphics[clip,width=0.5\textwidth]{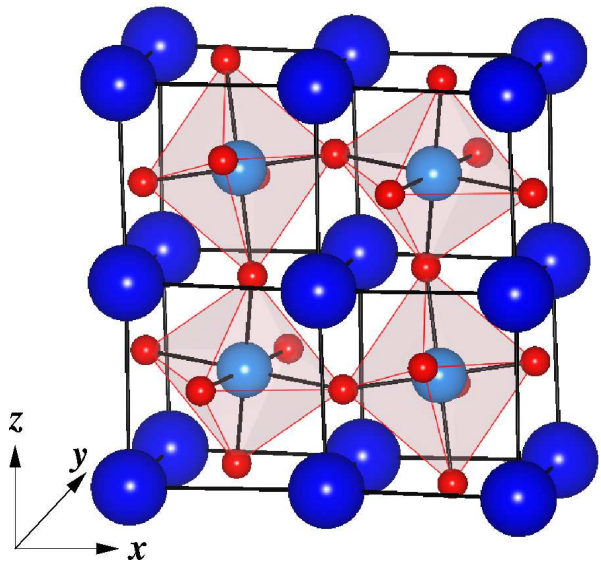}
  \caption{Representation of the JT/GFO distorted LaMnO$_3$ structure.
    Small red and light blue spheres indicate oxygen and manganese
    atoms, respectively, whereas the larger spheres refer to the La
    atoms. Plot done using the VESTA visualization
    program~\cite{vesta}.}
  \label{fig:0}
\end{figure}

Perovskite transition-metal oxides challenge electronic structure
theory since several decades, due to the variety of collective
structural, electronic, and magnetic phenomena which are responsible
for the formation of complex orbital- and spin-ordered
states~\cite{imada98,salamon01,dagotto}. A prototypical textbook
example of this class of materials is the antiferromagnetic insulator
LaMnO$_3$. The ground state electronic structure of LaMnO$_3$ is
characterized by the crystal-field induced breaking of the degeneracy
of the Mn$^{3+}$ 3$d^4$ manifold in the high-spin configuration
($t_{2g}$)$^3$($e_g$)$^1$, with the $t_{2g}$ orbitals lying lower in
energy than the two-fold degenerate $e_g$ ones. Due to the strong
Hund's rule coupling, the spins of the fully occupied majority
$t_{2g}$ orbitals are aligned parallel with the spin of the singly
occupied majority $e_g$ states on the same site. The orbital
degeneracy in the $e_g$ channel is further lifted via cooperative
Jahn-Teller (JT)
distortions~\cite{rodriguez98,chatterji03,sanchez03,qiu05}, manifested
by long and short Mn-O octahedral bonds alternating along the
conventional orthorhombic basal plane, which are accompanied by
GdFeO$_3$-type (GFO) checkerboard tilting and rotations of the oxygen
octahedra~\cite{elemans71,norby95,woodward} (see \fref{fig:0}).  As a
result, the ideal cubic perovskite structure is strongly distorted
into an orthorhombic structure with $Pbnm$
symmetry~\cite{elemans71,norby95}, and a $d$-type orbital-ordered (OO)
state emerges~\cite{murakami98}.  The corresponding occupied $e_g$
orbital can be written as $|\theta\rangle =
\mathrm{cos}\frac{\theta}{2}|3z^2-r^2\rangle +
\mathrm{sin}\frac{\theta}{2}|x^2 -
y^2\rangle~$\cite{kanamori60,yin06,pavarini10,sikora02}, with the sign
of $\theta\sim 108^\circ$ alternating along $x$ and $y$ and repeating
along $z$. This particular orbital ordering is responsible for the
observed A-type antiferromagnetic arrangement below
$T_{\mathrm{N}}=140$~K~\cite{wollan55,elemans71}. It was found that
long-range order disappears above 750~K, whereas a local JT distortion 
(without long-range order) remains (dynamically) active up to $>$ 1150~K~\cite{sanchez03,qiu05,pavarini10}.

The question of whether the origin of orbital ordering should be
attributed to a superexchange mechanism (O-mediated virtual hopping of
electrons between nearest neighbor $S=2$ Mn cations, associated with a
local Coulomb electron-electron interaction: $d^4_id^4_j
\rightleftharpoons d^3_id^5_j$)~\cite{kugel73} or to an
electron-lattice coupling effect (structural-induced splitting of the
degenerate $e_g$ levels)~\cite{kanamori60} has been the subject of
numerous
studies~\cite{feiner99,hotta00,bala00,ahn00,okamoto02,tyer04,zenia05,yin06,lin08,pavarini10}.
Considering that there is no clear experimental evidence to support
one mechanism over the other, the employment of theoretical models and
computer simulations has become an essential tool to explain the
complicated coupling between structural and electronic degrees of
freedom and to interpret the experimental observations.  On the basis
of model calculations, it has been recognized that the simultaneous
inclusion of both superexchange and JT interactions is crucial to
provide, to some extent, a satisfactory description of the observed
transition temperatures $T_{\mathrm{N}}$, $T_{\mathrm{OO}}$ and
$T_{\mathrm{JT}}$~\cite{feiner99,yin06,pavarini10}. This approach
typically relies on a suitable mapping between a realistic band
structure calculated e.g. via density functional theory
(DFT)~\cite{dft} and an effective many-body Hamiltonian, which is
often achieved by downfolding the relevant bands and constructing a
localized Wannier
basis~\cite{pavarini05,yin06,solovyev08,kovacik10,kovacik11}.

The quality and characteristics of the Wannier representation
inevitably depend on the underlying Kohn-Sham states.  It is well
known that the mean-field-type one-particle description of the
electronic structure within the standard local density
(LDA)~\cite{dft} or generalized gradient (GGA)~\cite{pbe}
approximations to DFT is incapable to correctly describe exchange and
correlation effects in the so called \emph{strongly-correlated
materials}, resulting, among other failures 
\footnote{
We note that the
underestimation of the band gap and related failures are of course
also partly due to the intrinsic limitation of the Kohn-Sham
approach, which is not meant to describe quasi-particle excitations
correctly.
}
in much too small band gaps and magnetic moments~\cite{imada98}. For this reason, the
DFT-derived subset of orbitals is typically employed as reference for
the one-electron (i.e.  non-interacting) part of the effective
Hamiltonian, where all approximated contributions coming from LDA/GGA
exchange-correlation effects are subtracted in order to avoid
double-counting~\cite{czyzyk94}. For example, in the DFT+DMFT method
(combination of DFT and dynamical mean-field theory
(DMFT))~\cite{dmft}, the effective Hamiltonian can be written as
$\hat{H} = \hat{H}_{\mathrm{DFT}} - \hat{H}_{\mathrm{dc}} +
\hat{H}_{U}$, where $\hat{H}_{\mathrm{DFT}}$ is the Kohn-Sham
Hamiltonian, $\hat{H}_{\mathrm{dc}}$ accounts for the double-counting
correction, and $\hat{H}_{U}$ represents the Hubbard-like term which
describes the electronic interactions in the strongly correlated
bands. A critical issue of the DFT+DMFT approach is that a well
defined expression for the double-counting potential is not known and
several forms have been suggested~\cite{czyzyk94,petukhov03}. Karolak
and coworkers have recently addressed this issue by treating the
double-counting term as an adjustable parameter and suggested that the
best agreement with experiment is achieved by setting the
double-counting potential in the middle of the gap of the impurity
spectral function~\cite{karolak10}. Within this context, it is
therefore justified to construct effective Hamiltonians starting from
band structures obtained using different schemes, such as e.g.
LDA$+U$~\cite{yin06} or hybrid functionals~\cite{jacob08}, which
usually provide much better gaps for semiconducting materials than
conventional DFT approximations and could therefore represent a more
appropriate ``non-interacting'' reference for model calculations.

For practical purposes, the most suitable starting point to study the
physics of complex transition-metal oxides is probably the
tight-binding (TB) scheme, which relies on a proper representation of
the electronic structure using a localized basis
set~\cite{imada98,dagotto}. Some of the authors have recently shown
that maximally localized Wannier functions (MLWFs) can be used to
extract an effective TB description of the $e_g$ subspace in
LaMnO$_3$~\cite{kovacik10,kovacik11}. The calculated TB parameters can
then be used to construct a simplified TB Hamiltonian in the form that
is very often used for the description of manganites,
$\hat{H}_{\mathrm{TB}} = \hat{H}_{\mathrm{kin}} +
\hat{H}_{\mathrm{Hund}} + \hat{H}_{\mathrm{JT}} +
\hat{H}_{\mathrm{e-e}}$, which then provides a very accurate
representation of the underlying Kohn-Sham band structure.

Motivated by the reasons outlined above, here we calculate MLWFs for
LaMnO$_3$ using several different methods, including both the
conventional GGA scheme and the more sophisticated GGA$+U$, hybrid
functionals, and GW approaches. Besides providing a detailed
description of the electronic and magnetic properties of LaMnO$_3$ at
various levels of theory, we investigate how the corresponding
differences in the treatment of exchange-correlation effects influence
the specific features of the MLWFs and the TB parameters derived from
them.

--------------------------------------------------------------------     
\section{\label{sec:comp}Methodology and Computational Details}
--------------------------------------------------------------------   

\subsection{\label{ssec:comp-dft} DFT-based calculations}

All our calculations are based on DFT within the
Perdew-Burke-Ernzerhof~\cite{pbe} (PBE) approximation to the
exchange-correlation energy.  The one-particle Kohn-Sham orbitals are
computed within a plane-wave basis employing two different codes: (i)
the program PWscf in combination with ultrasoft pseudopotentials
included in the {\sc Quantum ESPRESSO} package~\cite{espresso}, and
(ii) the projector augmented wave~\cite{paw1,paw2} (PAW) based Vienna
ab initio simulation package (VASP)~\cite{vasp1,vasp2}. In particular,
the PWscf program is used to benchmark the implementation of the
VASP2WANNIER90 interface at PBE and PBE$+U$ level. Due to the well
known limitations of standard DFT in describing the electronic
structure of ``strongly-correlated'' compounds, three different
corrections to the PBE wavefunctions are adopted: (i) PBE$+U$:
inclusion of a repulsive on-site Coulomb interaction $U$ following the
recipe of Dudarev et al.~\cite{dudarev}; (ii) Hybrid functionals:
suitable mixing between density functional and Hartree-Fock
theory~\cite{becke93} within the scheme proposed by Heyd, Scuseria,
and Ernzerhof (HSE06, HSE hereafter) in which one quarter of the
short-ranged exchange-correlation PBE functional is replaced by one
quarter of the short-ranged part of Hartree-Fock
exchange~\cite{hse,marsman09}; (iii) GW: explicit evaluation of the
self-energy $\Sigma = iGW$ within a partially self-consistent GW$_0$ procedure
made up of self-consistent update of the eigenvalues in the Green's function G and 
a fixed screened exchange $W_0$, evaluated using PBE wavefunctions~\cite{gw,franchini10}.
In accordance with previous studies~\cite{gw,franchini10}, five iterations were sufficient
to obtain quasiparticle energies converged to about 0.05 eV.

These four methodologies (PBE, PBE+U, HSE and GW) differ in a few fundamental issues:
(i) PBE relies on an approximate treatment of exchange-correlation effects; 
(ii) PBE+U contains the same PBE approximate correlation, but takes into account orbital 
dependence (applied to the $d$ states of manganese) of the Coulomb and exchange 
interactions which is absent in the PBE; 
(iii) HSE includes a portion of non-local fully orbital dependent exact exchange and PBE correlation
(iv) In GW exchange and correlation contributions are directly computed from the self-energy.

An identical technical setup is adopted for VASP and PWscf
calculations. All ground state electronic and magnetic properties are
calculated for the experimental low temperature $Pbnm$ structure
reported in \cite{elemans71} using a regular $\Gamma$-centered
7$\times$7$\times$5 and 6$\times$6$\times$6 k-point mesh in PWscf and
VASP, respectively (reduced to 4$\times$4$\times$4 at the GW$_0$ level),
and a plane wave energy cutoff of 35~Ry ($\approx\!{476}$~eV) and
300~eV in PWscf and VASP, respectively. 
Spin-polarized calculations were performed within a collinear setup without the inclusion of 
spin-orbit effects.
Except where otherwise noted, all PBE and PBE$+U$ results discussed in the present work refer to
PWscf calculations whereas HSE and GW$_0$ results are obtained using VASP.
In both PWscf and VASP we include the Mn(3$s$), Mn(3$p$), La(5$s$),
and La(5$p$) semi-core states in the valence. In PWscf the
(unoccupied) La(4$f$) states are excluded from the ultrasoft
pseudopotential, whereas they are present in the corresponding VASP
PAW potential.\footnote{ In the construction of the MLWFs within VASP
  we have shifted the La(4$f$) states to higher energies through the
  application of a large $U=10$~eV in order to avoid the overlap
  between La(4$f$) and unoccupied Mn($e_g$) states, which would
  otherwise deteriorate the disentanglement procedure.  }

To obtain the model TB parameters we perform additional calculations
for a simplified crystal structure with the same unit cell volume as
the experimental $Pbnm$ structure, but which involves only the
staggered ($Q^x$-type) JT distortion and no GFO distortion and no
orthorhombic deformation of the lattice vectors ($Q^z=0$).
See~\cite{ederer07,kovacik10,kovacik11} for more details and an exact
definition of the different distortion modes. The amplitude of $Q^x$
is 0.199 and 0.184~\AA\ in the experimental $Pbnm$ and in the
simplified JT($Q^x$) structure, respectively, and the amplitude of
$Q^z$ in the experimental $Pbnm$ structure is -0.071~\AA\ .

\subsection{\label{ssec:comp-wf}Maximally localized Wannier functions}

A set of $N$ localized Wannier functions $\vert w_{n\bi{T}} \rangle$
corresponding to a group of $N$ bands that are described by
delocalized Bloch states $\vert\psi_{m\bi{k}}\rangle$ is defined by
the following transformation:
\begin{equation}\label{eq:mlwf}
  \vert{w_{n\bi{T}}}\rangle = \frac{V}{\left({2\pi}\right)^{3}}
  \int_{\mathrm{BZ}} \mathrm{d}\bi{k} \left[{\sum_{m=1}^{N}
      U_{mn}^{\left(\bi{k}\right)}
      \vert{\psi_{m\bi{k}}}\rangle} \right]
  \mathrm{e}^{-\mathrm{i}\bi{k}\cdot\bi{T}}
  \,, 
\end{equation}
where $\bi{T}$ is the lattice vector of the unit cell associated with
the Wannier function, $m$ is a band index, $\bi{k}$ is the wave-vector
of the Bloch function, and the integration is performed over the first
Brillouin zone (BZ) of the lattice. Different choices for the unitary
matrices $\uuline{U}^{(\bi{k})}$ lead to different Wannier functions,
which are thus not uniquely defined by \eref{eq:mlwf}. A unique set of
\emph{maximally localized Wannier functions} (MLWFs) can be generated
by minimizing the total quadratic spread of the Wannier
orbitals~\cite{1997_marzari}.

Once the transformation matrices $\uuline{U}^{(\bi{k})}$ are
determined, a TB representation of the Hamiltonian in the basis of
MLWFs is obtained:
\begin{equation}\label{eq:tbh}
  \hat{H} = \sum_{\bi{T}, \Delta\bi{T}} h_{nm}^{\Delta\bi{T}}
  \, \hat{c}^\dagger_{n\bi{T}+\Delta\bi{T}} \hat{c}_{m\bi{T}}
  \ + \mathrm{h.c.} \ ,
\end{equation}
with
\begin{equation}\label{eq:hr}
  h^\bi{T}_{nm} = \frac{V}{(2\pi)^3} \int_\mathrm{BZ} \mathrm{d}\bi{k}
  \left[
    \sum_{l} \left(U^{(\bi{k})}_{ln}\right)^* \epsilon_{l\bi{k}} \, U^{(\bi{k})}_{lm}
  \right]
  \mathrm{e}^{-\mathrm{i}\bi{k}\cdot\bi{T}}  \,.
\end{equation}
Here, $\epsilon_{l\bi{k}}$ is the eigenvalue corresponding to Bloch
function $\vert\psi_{l\bi{k}}\rangle$. For cases where the bands of
interest do not form an isolated set of bands but are entangled with
other bands, a two step procedure for obtaining the unitary
transformation matrices (which in this case are typically rectangular)
is employed~\cite{2001_souza}. We note that $\bi{T}$ and
$\Delta\bi{T}$ in \eref{eq:mlwf}-\eref{eq:hr} indicate lattice
translations, whereas for crystal structures with more than one atom
per unit cell, $n$ and $m$ generally represent combined orbital, spin,
and site indeces, specifying the various orbitals at all sites within
the primitive unit cell.

Based on the projected densities of states (PDOS) calculated within
DFT, we determine a suitable energy window for the construction of the
MLWFs (more details follow in \sref{ssec:res-mlwf}). MLWFs are
constructed by merging PWscf and VASP with the wannier90 code using
the available PW2WANNIER90 tool~\cite{wannier90,marzari12} and the newly
introduced VASP2WANNIER90 interface, respectively. Technical details 
on the construction of MLWFs within the PAW formalism can be found in 
Ref.\cite{Ferretti}.
Starting from an initial projection of the Bloch bands onto atomic $e_g$ basis
functions \mbox{$\vert{3z^2-r^2}\rangle$} and
\mbox{$\vert{x^2-y^2}\rangle$} centered at different Mn sites within
the unit cell, we obtain a set of two $e_{g}$-like MLWFs per spin
channel for each Mn site. The spread functional (both gauge-invariant
and non-gauge-invariant parts) is considered to be converged if the
corresponding fractional change between two successive iterations of
the spread minimization is smaller than $10^{-10}$.

{\em Practical instructions for the use of VASP2WANNIER90:} VASP uses
wannier90 in library mode to generate all ingredients which are
required to run the wannier90 code as a post-processing tool.  

Apart from the main wannier90 input file (wannier90.win) the input files
needed by wannier90 are~\cite{wannier90}: (i) the overlaps between the
cell periodic parts of the Bloch states (wannier90.mmn), (ii) the
projections of the Bloch states onto trial localized orbitals
(wannier90.amn), and (iii) the eigenvalues file (wannier90.eig). This
set of files is generated by VASP by setting LWANNIER90 = .TRUE. in
the main VASP input file (INCAR). If the file wannier90.win already
exists, VASP will properly generate the files (i)-(iii) according to
the instructions specified in wannier90.win. If wannier90.win does not
exist, VASP will generate a default wannier90.win file, which should
be suitably modified in accordance to the keyword list described in
the wannier90 user guide~\cite{wannier90online} in order to tell VASP
what quantities to compute. Then, VASP has to be run again in order to
create the additional wannier90 input files. To construct the UNK
files (the periodic part of the Bloch states represented on a regular
real space grid), which are required to plot the MLWFs, it is
necessary to set LWRITE\_UNK = .TRUE. in the INCAR file. In a
spin-polarized calculation two sets of input files are generated 
(VASP2WANNIER90 is employed only once to generate the files wannier90.mmn,
wannier90.amn, and wannier90.eig. These files are then used as input files 
for wannier90, which is serially run for each energy window).  
Please refer to the online documentation of wannier90 for a detailed
description of all relevant instructions~\cite{wannier90online}.

------------------------------------------------------------------------------------------------     
\section{\label{sec:res}Results and Discussion}
------------------------------------------------------------------------------------------------    

In this section we will first present and compare the electronic and
magnetic ground state obtained within the various levels of
approximation (PBE, PBE$+U$, HSE and GW$_0$), before we will describe the
downfolding of the resulting band structure by Wannier function
decomposition. Finally, TB parameterizations corresponding to
effective $e_g$ models, either with or without explicit
electron-electron interaction term, will be derived from these
results, and implications of the different underlying band-structures
will be discussed.

\subsection{\label{ssec:res-em} Electronic and magnetic ground state}

\begin{figure}
  \includegraphics[clip,width=1.0\textwidth]{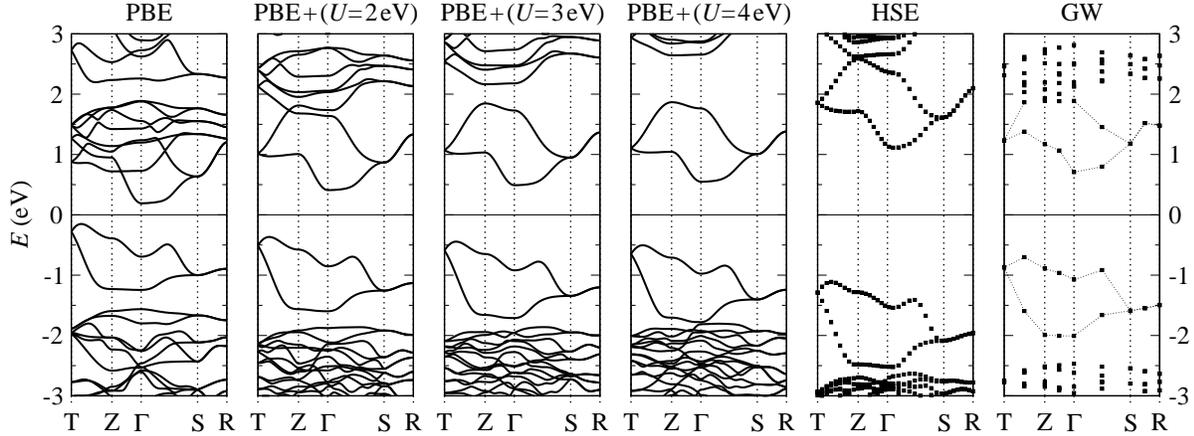}
  \caption{Calculated band structure along certain high-symmetry
    directions within the BZ. Each panel reports results obtained by a
    different method, as specified in the panel title. E=0 is aligned to the middle of the gap.}
  \label{fig:1}
\end{figure}

\begin{figure}
  \centering
  \includegraphics[clip,width=0.5\textwidth]{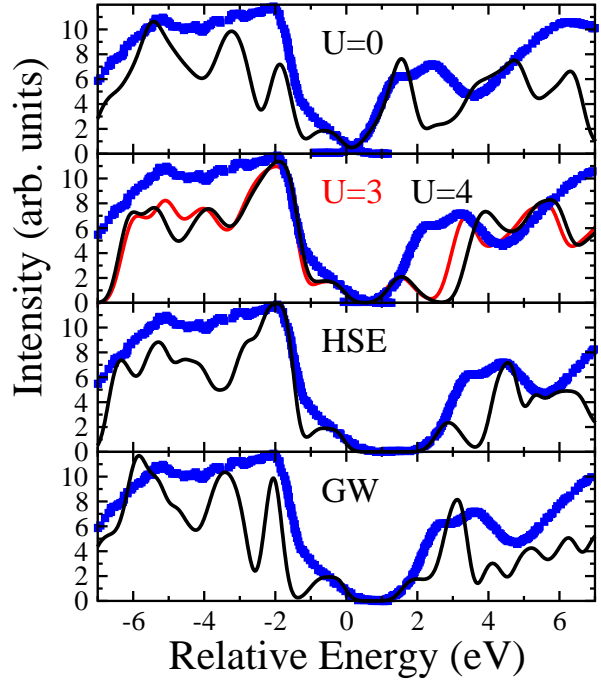}
  \caption{Comparison between experimental~\cite{park96} (blue
    squares) and calculated valence and conduction band spectra for
    PBE, PBE$+U$ ($U=3$ and 4~eV), HSE, and GW$_0$. The calculated and
    measured spectra have been aligned by overlapping the valence band
    maxima and conduction band minima.}
  \label{fig:2}
\end{figure}

The calculated band structures are displayed in \fref{fig:1} and the
corresponding indirect ($E_\mathrm{i}$) and smallest direct
($E_\mathrm{d}$) band gaps are listed in \tref{tab:1}. The calculated
valence and conduction band spectra and the PDOS (corresponding to
Mn($e_g$), Mn($t_{2g}$), and O($p$) states), are represented in
\fref{fig:2} and \fref{fig:bs-mlwf}, respectively.

It can be seen from \fref{fig:1} that the eigenvalue dispersion in
LaMnO$_3$ is characterized by an insulating state with indirect energy
gap. By comparing with the PDOS shown in \fref{fig:bs-mlwf}, it
becomes clear that within all methods the Mott-Hubbard gap is opened between
occupied and empty states with predominant Mn($e_g$) character. 
While the width of the band gap differs strongly between the various
methods, each one is in good agreement with previous
LDA/GGA~\cite{pickett97,sawada96,Ravindran02,hashimoto},
(LDA/GGA)$+U$~\cite{sawada96,hashimoto}, and hybrid
functionals~\cite{munoz04}, respectively (see \tref{tab:1}). 
Our partially-self consistent GW$_0$ cannot be 
directly compared with the single-shot G$_0$W$_0$ results of Nohara {\em et al.}\cite{nohara06} since the latter 
depend much more on the initial LDA wavefunction and consequentially convey a smaller bandgap.

\begin{table}
  \caption{Collection of calculated (present work and previous
    studies) and experimental value for the indirect
    ($E_{\mathrm{i}}$) and direct ($E_{\mathrm{d}}$) band gap of
    LaMnO$_3$.  The measured values refer to optical
    conductivity~\cite{arima,Jung97,Jung98}, Raman~\cite{Kruger04}, and photoemission~\cite{saitoh}
    experiments.}
  \begin{indented}
  \item[]
    \begin{tabular}{@{}ccccccccc}
      \br
      \multicolumn{7}{c}{This Work} \\
      &HSE  &  GW$_0$@PBE &  PBE      &\multicolumn{3}{c}{PBE$+U$} \\
      &&         &           &$U=2$  &$U=3$  &$U=4$  \\
      \mr
      $E_{\mathrm{i}}$        &2.25 & 1.41    &      0.38 & 0.82 & 0.98 & 1.10 \\
      $E_{\mathrm{d}}        $&2.55 & 1.68    &      0.75 & 1.15 & 1.30 & 1.42 \\
      \br
      \multicolumn{7}{c}{Previous studies} \\
      &B3LYP\cite{munoz04} & G$_0$W$_0$@LDA\cite{nohara06} & GGA\cite{hashimoto} & GGA$+U$\cite{hashimoto} &
      \multicolumn{2}{c}{Expt.}\\
      &&         &           &$U$=2  &     &      \\
      \mr
      $E_{\mathrm{i}}$&2.3   & 0.82   & 0.27    & 0.81   &      &     \\
      $E_{\mathrm{d}}$&      & 1.00   & 0.70    & 1.18   &  \multicolumn{2}{c}{1.1$^a$, 1.9$^b$, 2.0$^{c,d}$, 1.7$^e$} \\
      \br
    \end{tabular}
  \end{indented}
  \label{tab:1}
\begin{flushleft}
$^a$Ref. \cite{arima}, $^b$Ref. \cite{Jung97}, $^c$Ref. \cite{Jung98}, $^d$Ref. \cite{Kruger04}, $^e$Ref. \cite{saitoh},
\end{flushleft}
\end{table}

Due to the inadequate treatment of exchange-correlation effects,
conventional PBE-DFT leads to a significant underestimation of
$E_{\mathrm{d}}^{\mathrm{PBE}}=0.75$~eV compared to the experimental
values obtained from optical conductivity measurements
(1.1~eV~\cite{arima}, 1.9~eV~\cite{Jung97}, 2.0~eV~\cite{Jung98}), Raman (2.0~eV~\cite{Kruger04}),
 and photoemission data (1.7~eV~\cite{saitoh}).
In addition, the uppermost filled Mn($e_g$) bands (with energies in
the region between $-$1.3~eV and 0.0~eV) are well separated from the
lower-lying mostly Mn($t_{2g}$)- and O($p$)-like states (below
$-1.5$~eV). In contrast, while the lower part of the group of bands
immediately above the gap (up to about 2~eV) exhibits predominant
local majority spin $e_g$ character, these bands are strongly
entangled with local minority spin $t_{2g}$ states at slightly higher
energies (between approximately 1-2~eV). The inclusion of the on-site
interaction term within the PBE$+U$ approach, separates these
higher-lying local minority spin $t_{2g}$ states from the local
majority $e_g$ bands directly above the gap for $U>2$~eV. Furthermore,
increasing $U$ also increases the band gap
($E_{\mathrm{d}}^{\mathrm{PBE}+U}=1.42$~eV for $U=4$~eV) and lowers
the filled $e_g$ states relative to the bands with dominant
Mn($t_{2g}$) and O($p$) character, which leads to an appreciable
overlap between these sets of bands around the $\Gamma$ point for
$U=4$~eV.

Changing to a more elaborate treatment of the exchange-correlation
kernel, we observe that HSE provides a value of the bandgap
($E_{\mathrm{d}}^{\mathrm{HSE}}=2.55$~eV) that is significantly larger (by $\approx$ 0.5 eV)
than the experimental measurements. This is in line with previous
hybrid functional estimates based on the B3LYP approach implemented
within a Gaussian basis set~\cite{munoz04}. By comparing the PBE and
HSE band gap one could argue that a smaller portion of exact
Hartree-Fock exchange should be included in the hybrid functional
framework in order to obtain a better agreement with experiment.
Indeed, a reduced mixing parameter $a_{\mathrm{mix}}=0.15$ shrinks the
direct gap down to 1.79~eV, almost on par with the photoemission
measurements of Saitoh and coworkers~\cite{saitoh}, 
and with the more recent optical conductivity data of Jung {\em et al.}\cite{Jung97,Jung98}, 
and Kr\"uger {\em et al.}\cite{Kruger04}.
LaMnO$_3$ therefore seems to represent another example for which the one-quarter
compromise (mixing 1/4 of exact exchange with 3/4 of DFT exchange) is
not the ideal choice~\cite{Franchini11}. 
Finally, the parameter-free GW$_0$ technique leads
to a quite satisfactory prediction of the band gap, $E_{\mathrm{d}}^{\mathrm{GW_0}}=1.68$~eV, 
and about significantly larger than the only
previous single-shot (i.e. perturbative) G$_0$W$_0$ study of Nohara {\em et al.} based on 
initial LDA wavefunctions~\cite{nohara06}.  Similarly to HSE
and PBE$+U$ (for $U=3$~eV), GW$_0$ deliver $e_g$ bands around $\rm E_F$
well separated from the O($p$) and Mn($t_{2g}$)
bands below and, to a lesser extent, above (there is an appreciable mixing of Mn($e_{g}$) and Mn($t_{2g}$) states 
along the T-Z-$\Gamma$ path around 2 eV), in clear contrast with the PBE picture which predicts a certain degree of
overlap between the $e_g$ bands and the higher lying $t_{2g}$ bands.

In order to provide further assessment of the quality of the various
methods in describing the electronic structure of LaMnO$_3$ we compare
in \fref{fig:2} the simulated valence and conduction band spectra with
the corresponding photoemission spectroscopy and X-ray absorption
spectroscopy data~\cite{park96}. For negative energies (occupied
states) none of the four methods differs dramatically from the
experimental spectrum, even though the multi-peak structures in the
range of $-$7~eV to $-$4~eV seen within PBE$+U$ and HSE do not have a
clear experimental correspondence, whereas PBE and GW$_0$ profiles 
better follow the main three experimental peak/shoulders. 
The situation is more critical for the unoccupied region,
since none of the methods is capable to correctly reproduce the two-peaks
structure characterizing the onset of the conduction band right above $\rm E_F$.
These two peaks could be interpreted as formed by $e_g$ (lower one) and $t_{2g}$  
(second ones) contributions and are described differently by the various schemes, 
following the corresponding band dispersions discussed in Fig. \ref{fig:1}:
(i) PBE both peaks merge in one single strong electronic signal, reflecting the 
large overlap between $e_g$ and $t_{2g}$ bands right above $\rm E_F$;
(ii) in PBE+U the two peak are much too separated, reflecting the wide
$e_g$-$t_{2g}$ band splitting;
(iii) HSE and GW$_0$ are rather similar.
Their spectra are characterized by a lower $e_g$ small bunch of states (onset of the conduction band spectra)
associated to a more intense $t_{2g}$-like peak, but the GW$_0$ 
$e_g$/$t_{2g}$ splitting ($\approx$ 1.4 eV) better matches the experimental one
($\approx$ 1.1 eV) as compared to the larger HSE splitting ($\approx$ 1.7 eV).
From these results we can infer that GW$_0$ and HSE convey
the most satisfactory picture in terms of peak position and
corresponding spectral weight for both occupied and unoccupied states,
with GW$_0$ better reproducing the splitting between the two lower conduction peaks.
However, it should be noted that the relative weights of the two lower conduction peaks 
do not match with experiment, indicating that it is necessary go beyond the GW approximation
to obtain a refined agreement with experiment, for instance using the Bethe-Salpeter equation
(this is beyond the scope of the present study).
We underline once more that unlike PBE+U and HSE (in which the proper adjustment 
of the parameters $U$ and $a_{\mathrm{mix}}$ can cure the bandgap problem
and lead to values of the gap close to the experimental ones), the parameter-free
G$W_0$ scheme is capable to provide a rather accurate picture without the
need of any adjustable parameter.

Next, we analyze the magnetic properties in terms of the
nearest-neighbor magnetic exchange interactions within the
orthorhombic $ab$ plane ($J_{ab}$) and along $c$
($J_c$)~\cite{solovyev96,munoz04,evarestov}. This will provide further
insights into the performance of the various methods with respect to
energetic properties of LaMnO$_3$. By mapping the calculated total
energies for different magnetic configurations onto a classical
Heisenberg Hamiltonian
$H=-\frac{1}{2}\sum_{i{\neq}j}J_{ij}\,{{S_i\cdot S_j}}$, the following
equations for $J_{ab}$ and $J_c$ can be obtained (see also
\cite{munoz04,evarestov}):
\begin{equation}
  E_{\mathrm{FM}} - E_{\mathrm{AAF}} = -32 J_c 
\end{equation}
\begin{equation}
  E_{\mathrm{CAF}} - E_{\mathrm{FM}} = 64 J_{ab} \,.
\end{equation}
Here, $E_{\mathrm{FM}}$ corresponds to the total energy for the
ferromagnetic (FM) configuration, whereas $E_{\mathrm{AAF}}$ and
$E_{\mathrm{CAF}}$ indicate the total energies associated with
antiferromagnetic (AFM) ordering along $z$, and a two-dimensional
checker-board like arrangement within the $xy$ plane,
respectively~\cite{munoz04}.

The values of $J_{ab}$ and $J_c$ obtained using the various methods
considered within this work are listed in \tref{tab:2} along with the
calculated magnetic moments at the Mn site. We note that, due to the
neglect of orbital degrees of freedom which in LaMnO$_3$ are strongly
coupled to spin degrees of freedom, it is not obvious whether a
classical Heisenberg model is well suited to give a complete picture
of the magnetic properties of LaMnO$_3$. Nevertheless it can still
provide an accurate parameterization of the energy differences between
the various magnetic configurations.  However, the quantitative
comparison with the experimental coupling constants derived from
spin-wave spectra, i.e. small fluctuations around the AFM ground
state, should be taken with care. In view of this, we can draw the
following conclusions about the efficiency of the various DFT and
beyond-DFT methods employed in the present study: (i) the magnetic
energy differences exhibit appreciable variation between VASP and
PWscf leading to differences of about 1-2 meV in the magnetic coupling
constants. This is most likely due to the different pseudopotential
technique employed in the two codes (PAW method vs.  ultrasoft
pseudopotential), which lead to qualitative differences especially at
PBE$+U$ level, as discussed below.  
A more elaborate discussion on the performance of different functionals and methods
in predicting the magnetic couplings is given in Refs. \cite{hashimoto, evarestov}, where
it is concluded that the PAW values are very similar to the full potential FLAPW ones.
(ii) In both codes PBE gives the
correct A-AFM ground state, delivering a negative $J_c$
($J_c^\mathrm{VASP} = -2.13$~meV, $J_c^\mathrm{PWscf}=-0.81$~meV) and
a positive $J_{ab}$ ($J_{ab}^\mathrm{VASP}=3.22$~meV,
$J_{ab}^\mathrm{PWscf}=4.56$~meV). (iii) The ``$+U$'' correction to
PBE decreases the $E_{\mathrm{FM}} - E_{\mathrm{AAF}}$ energy
difference and eventually leads to the prediction of a FM ground state
for $U$ larger than a certain value. This critical value is rather
different within the two codes used in this study: $J_c$ becomes
positive for $U=2$~eV and $U=4$~eV, in PWscf and VASP, respectively.
We note that this difference is almost entirely due to the difference
in the corresponding PBE results. The $U$-induced changes in the
magnetic coupling constant $J_c$ relative to the $U=0$ reference are
nearly identical within the two codes. (iv) While $J_{c}$ within HSE
and PBE+$U$(VASP) are very similar for $U$ between 2-3~eV, the ratio
between $J_c$ and $J_{ab}$ is rather different within the two
approaches. (v) Within the limitations regarding the applicability of
a Heisenberg picture to LaMnO$_3$ stated above, HSE seems to be most
consistent with the values of the magnetic coupling constants derived
from neutron diffraction measurements of spin-wave
spectra~\cite{moussa96} and magnon data~\cite{hirota96}.  This further
confirms the predictive power of HSE in describing exchange
interactions in transition metal oxides, as compared to other
available beyond-DFT schemes~\cite{archer}.

We can also see that all methods result in values for the local
magnetic moments of the Mn cation that are within the range of
variation of the experimental data. Generally, increasing $U$ within
PBE+$U$ leads to a more localized magnetization density compared to
PBE, and thus increases the local magnetic moments.

On the basis of the above analysis both of the electronic and magnetic
properties of LaMnO$_3$, we can conclude that HSE and, when applicable, 
GW$_0$ (the calculation of magnetic energies at GW level to extract exchange coupling 
constants is presently not possible, or at least extremely difficult) are the most
consistent with the available experimental data in terms of spectral
properties, electronic structure and magnetic exchange interactions of LaMnO$_3$. 
In view of this, we can now proceed to the discussion of the Wannier-based description 
of the $e_g$ bands and the associated TB parameterization.

\begin{table}
  \caption{PBE, PBE$+U$, HSE and GW$_0$ derived magnetic exchange
    parameters (meV) and magnetic moment at Mn sites $\mu$
    ($\mu_{\mathrm{B}}$).  The experimental and previously published
    computed data are taken from: $^{a}$~Ref.~\cite{hashimoto},
    $^{b}$~Ref.~\cite{munoz04}, $^{c}$~Ref.~\cite{moussa96},
    $^{d}$~Ref.~\cite{elemans71},   
    $^{e}$~Ref.~\cite{hauback96}, and
    $^{f}$~Ref.~\cite{hirota96}
    }
  \begin{indented}
  \item[]
    \begin{tabular}{@{}lccc}
      \br
      & $J_{ab}$       &  $J_c$      & $\mu$ \\
      \mr
      \multicolumn{4}{c}{PWscf} \\
      PBE                      & 4.56        & $-$0.81     &  3.67   \\ 
      $U=2$~eV                 & 5.02        &    0.37     &  3.82   \\ 
      $U=3$~eV                 & 5.30        &    0.98     &  3.89   \\ 
      $U=4$~eV                 & 5.63        &    1.55     &  3.96   \\ 
      \multicolumn{4}{c}{VASP} \\
      PBE                      & 3.22        & $-$2.13     &  3.50   \\
      $U=2$~eV                 & 3.54        & $-$0.84     &  3.68   \\
      $U=3$~eV                 & 3.57        & $-$0.30     &  3.76   \\
      $U=4$~eV                 & 3.61        &    0.17     &  3.83   \\
      HSE                      & 2.56        & $-$0.53     & 3.74  \\ 
      GW$_0$                   &             &             & 3.51  \\
      \multicolumn{4}{c}{Previous studies} \\
      GGA$+U$ ($U=2$~eV)$^{a}$ &             & $-$1.30                    & 3.46  \\ 
      B3LYP$^{b}$              & 2.09        & $-$1.01                  & 3.80  \\ 
      Expt                            & 1.66$^{c}$  & $-$1.16$^{c}$ & 3.87$^{c}$, 3.7$\pm$0.1$^{d}$, 3.4$^{e}$ \\ 
                                      & 1.67$^{f}$  & $-$1.21$^{f}$ &  \\
      \br
    \end{tabular}
  \end{indented}
  \label{tab:2}
\end{table}

\subsection{\label{ssec:res-mlwf} Maximally localized Wannier
  functions}
\label{ssec:mlwf}

In this section we present the details for the construction of the
MLWFs with predominant $e_{g}$ character from the calculated bands
around the gap. In a TB picture, these MLWFs can be seen as
``antibonding'' bands resulting from the $\sigma$-type hybridization
between the Mn($d$) and O($p$) atomic orbitals. Note that in this and
the next section the discussion of the PBE$+U$ results refers to the
representative value of $U=3$~eV, unless explicitly stated otherwise.

\begin{figure}
  \includegraphics[clip,width=1.0\textwidth]{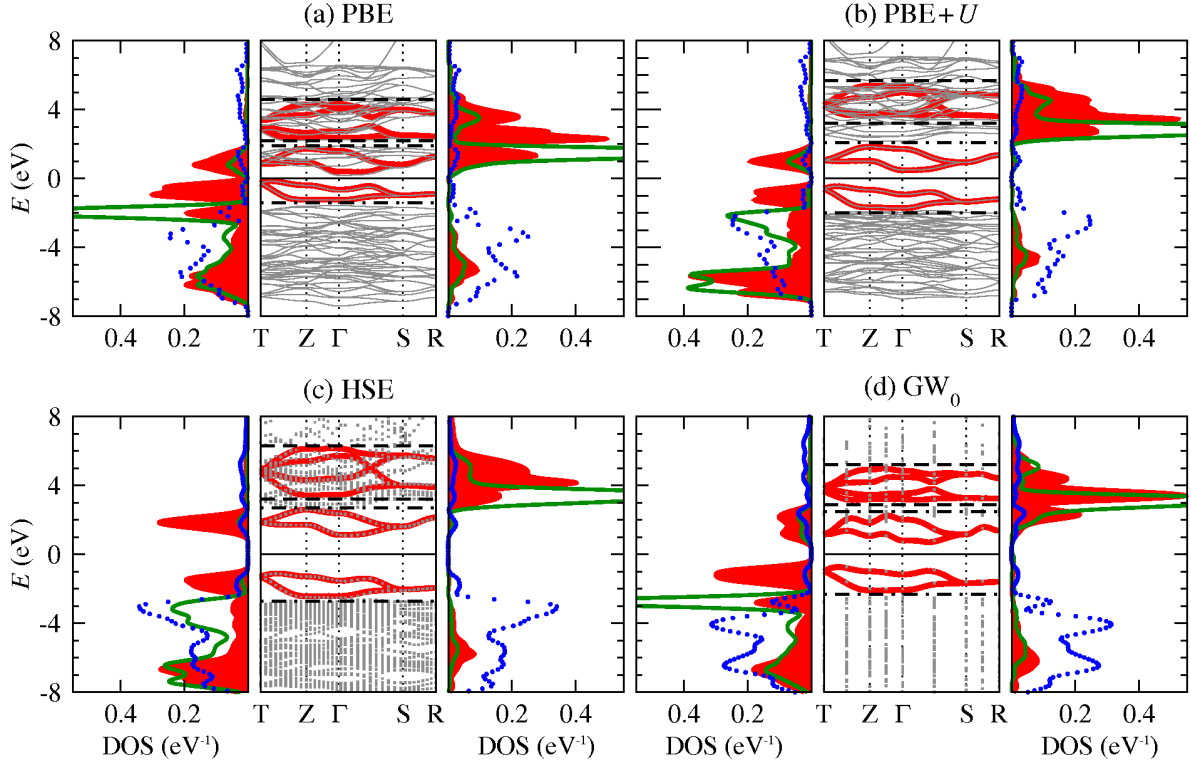}     
  \caption{Effective $e_g$ MLWF bands (thick red lines) for LaMnO$_3$
    superimposed to the {\em ab initio} electronic bands (gray thin
    solid/dotted lines) and associated normalized PDOS (to the left
    and right of the band structure plots) corresponding to Mn($e_g$)
    (red filled areas), Mn($t_{2g}$) (green lines), and O($p$)
    character (blue dots). In the left/right PDOS graphs, Mn($d$)
    PDOSs correspond to the local majority/minority Mn sites while the
    O($p$) PDOS is calculated as an average over all O sites. 
    The two energy windows used in the wannier-downfolding are indicated by
    dashed and dot-dashed lines. The Fermi level (E=0 eV) is set in the 
    middle of the gap.}
  \label{fig:bs-mlwf}
\end{figure}

\Fref{fig:bs-mlwf} shows the PBE and beyond-PBE (PBE$+U$, HSE and GW$_0$)
band structures and the corresponding PDOS with Mn($e_{g}$),
Mn($t_{2g}$), and O($p$) character. Apart from the obvious
hybridization between Mn($d$) and O($p$) states, ``$e_g$-like''
orbitals at a certain site can hybridize with ``$t_{2g}$-like''
orbitals at a neighboring site as a result of the tilt and rotation of
the oxygen octahedra. This leads to bands with mixed $e_{g}$/$t_{2g}$
character (note the bands around the gap with strong PDOS components
of both $e_{g}$ and $t_{2g}$ character). Due to this strong mixing it
is not possible to construct 8 $e_{g}$ character MLWFs within one
energy window used in the disentanglement procedure.\footnote{In the
  antiferromagnetic case each band is of course two-fold degenerate
  with respect to the global spin projection. Here and in the
  following we refer to such pairs of spin-degenerate bands as ``one
  band''.} The corresponding energy window would inevitably also
contain the local minority spin ``$t_{2g}$'' bands. Since due to the
GFO distortion these bands can hybridize with the minority spin
``$e_g$'' bands, this would lead to MLWFs with strongly mixed
$e_g$/$t_{2g}$ character. To circumvent this problem, we therefore
construct two separate sets of 4 local majority and 4 local minority
spin MLWFs using two different energy windows~\cite{kovacik10}. These
energy windows have to be chosen carefully for each individual method.
(This problem is not present for the purely JT($Q^x$) distorted
structure, from which we derive most of the model parameters, see
\sref{ssec:res-tb}. In this case we calculate a full set of 8 MLWFs).

To find a suitable energy window is quite straightforward for the
local majority spin case. The upper bound of the energy window is
determined by the upper bound of the highest (in energy) peak of the
local majority spin Mn($e_{g}$) PDOS, while the lower bound of the
energy window should be placed above the occupied bands with strong
O($p$) and/or local majority spin Mn($t_{2g}$) character. It can be
seen from \fref{fig:bs-mlwf} (and perhaps more clearly from
\fref{fig:1}), that both the lower and the upper bound fall within
small gaps separating the bands within the energy window from other
bands at lower and higher energies. 
Furthermore, for PBE+U and HSE the MLWFs can be constructed
from a completely isolated set of bands, whereas in the case of PBE and GW$_0$
additional bands with predominant minority spin Mn($t_{2g}$) character
are included in the energy window.
However, due to the different local spin projection, these
latter bands have no noticeable effect on the final MLWFs.

For the local minority spin MLWFs, the upper bound of the energy
window can be found in the same way as for the local majority spin
bands. Within PBE the lower bound is also easily determined,
since it falls within a small gap separating the local minority spin
bands with predominant $e_{g}$ and $t_{2g}$ character. However, no
such gap exists within PBE$+U$, HSE, and GW$_0$, and it is thus not possible to
fully exclude the $t_{2g}$ character from the resulting MLWFs.
Instead, the lower bound of the energy window has to be carefully
adjusted by manually checking the $e_g$ character of the calculated
MLWFs in real space.

The band dispersion of the so-obtained MLWFs is shown in
\fref{fig:bs-mlwf} as thick red lines. The 4 (energetically lower)
local majority MLWF bands follow very closely the underlying
PWscf/VASP bands and the overall dispersion is very similar for all
methods.  Despite the strong band-entanglement, the dispersion of the
4 (energetically higher) local minority MLWF bands is also very
similar within all methods. Only the energetically lowest local
minority spin band within PBE$+U$ and HSE exhibits strong deviations
from the corresponding PBE and GW$_0$ case. This is due to the
above-mentioned difficulty to exclude the $t_{2g}$ character in a
controlled way. Conclusions drawn from such sets of MLWFs should
therefore be taken with care. Overall, we note that the similarities
in the band structure and PDOS between PBE and GW$_0$ as well as between
PBE$+U$ and HSE, regarding the degree of hybridization between
Mn($e_g$), Mn($t_{2g}$) and O($p$) orbitals, that have been pointed
out in the previous section, are also reflected in the MLWF bands.

\begin{figure}
  \centering
  \includegraphics[clip,width=0.75\textwidth]{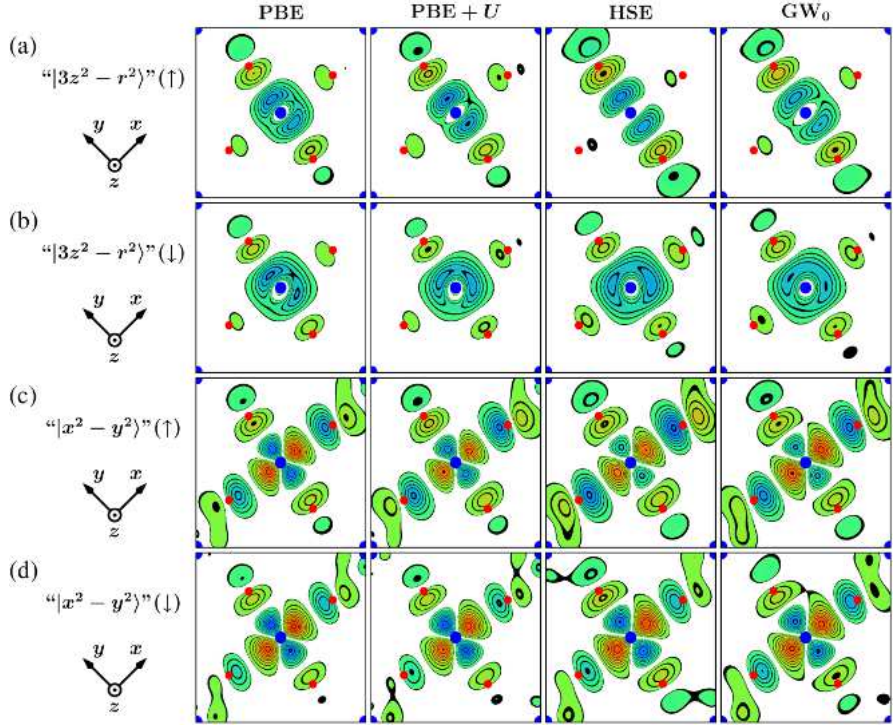}
  \caption{Real space representation of the four $e_g$ MLWFs
    corresponding to a certain Mn site, projected on the $xy$ plane
    cutting through the Mn site. Black iso-lines correspond to $\pm
    N/\sqrt{V}$ with integer $N \ge 1$, the white region is defined by
    values in the interval $[-1/\sqrt{V},+1/\sqrt{V}]$, where $V$ is
    the volume of the unit cell. Blueish/reddish hue denotes
    negative/positive values of MLWFs and Mn and O atoms are shown as
    blue and red spheres, respectively.}
  \label{fig:rs-mlwf}
\end{figure}
\begin{figure}
  \centering
  \includegraphics[clip,width=1.00\textwidth]{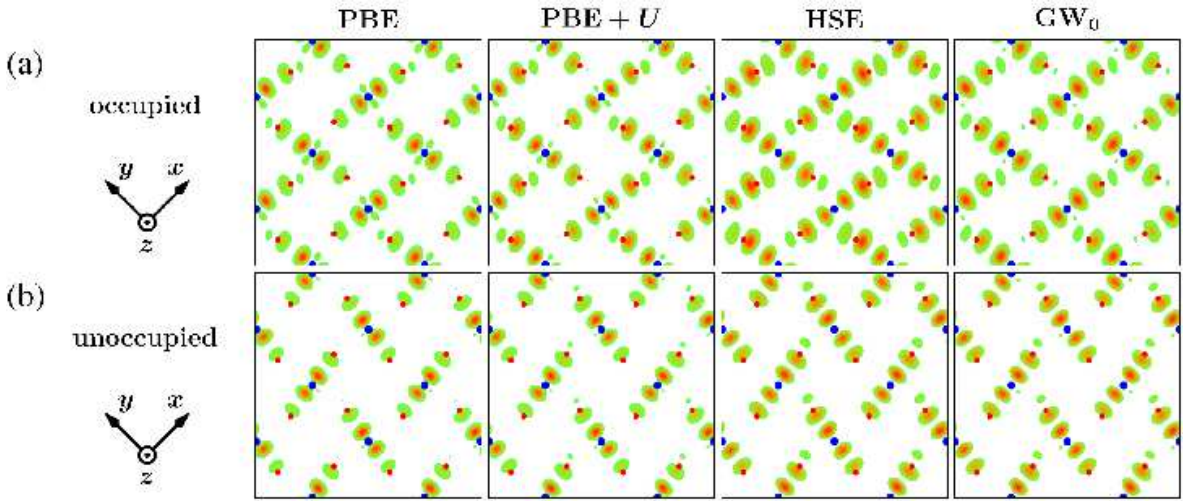}
  \caption{Charge density isosurfaces of the orbitally ordered states associated to the highest 
occupied (a) and lower unoccupied (b) MLWFs orbitals.
Color coding and symbols are the same as in Fig. \ref{fig:rs-mlwf}.}                            
  \label{fig:OOc-mlwf}
\end{figure}
\begin{figure}
\centering
\includegraphics[clip,width=0.22\textwidth]{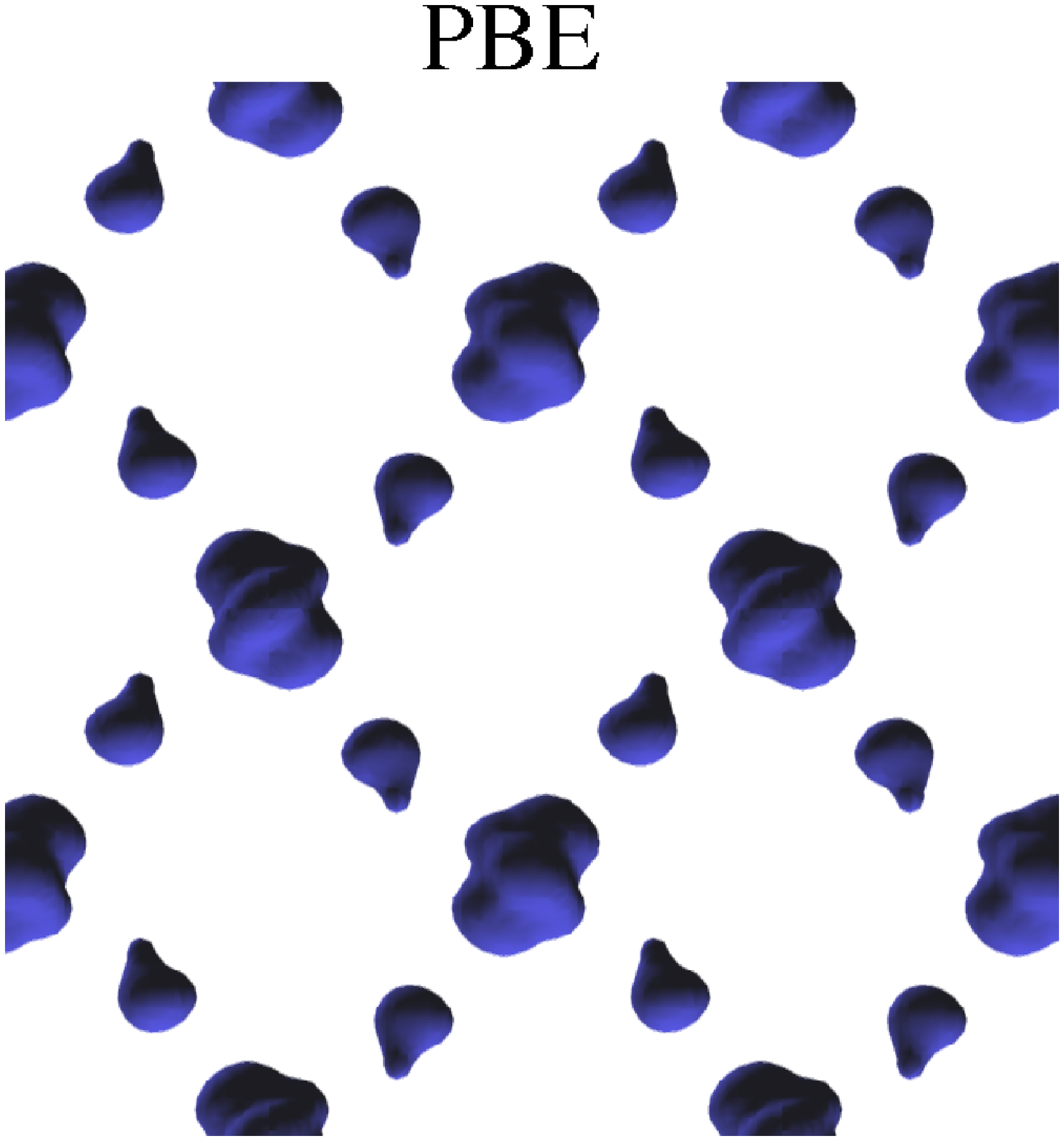}
\hspace{0.1cm}
\includegraphics[clip,width=0.22\textwidth]{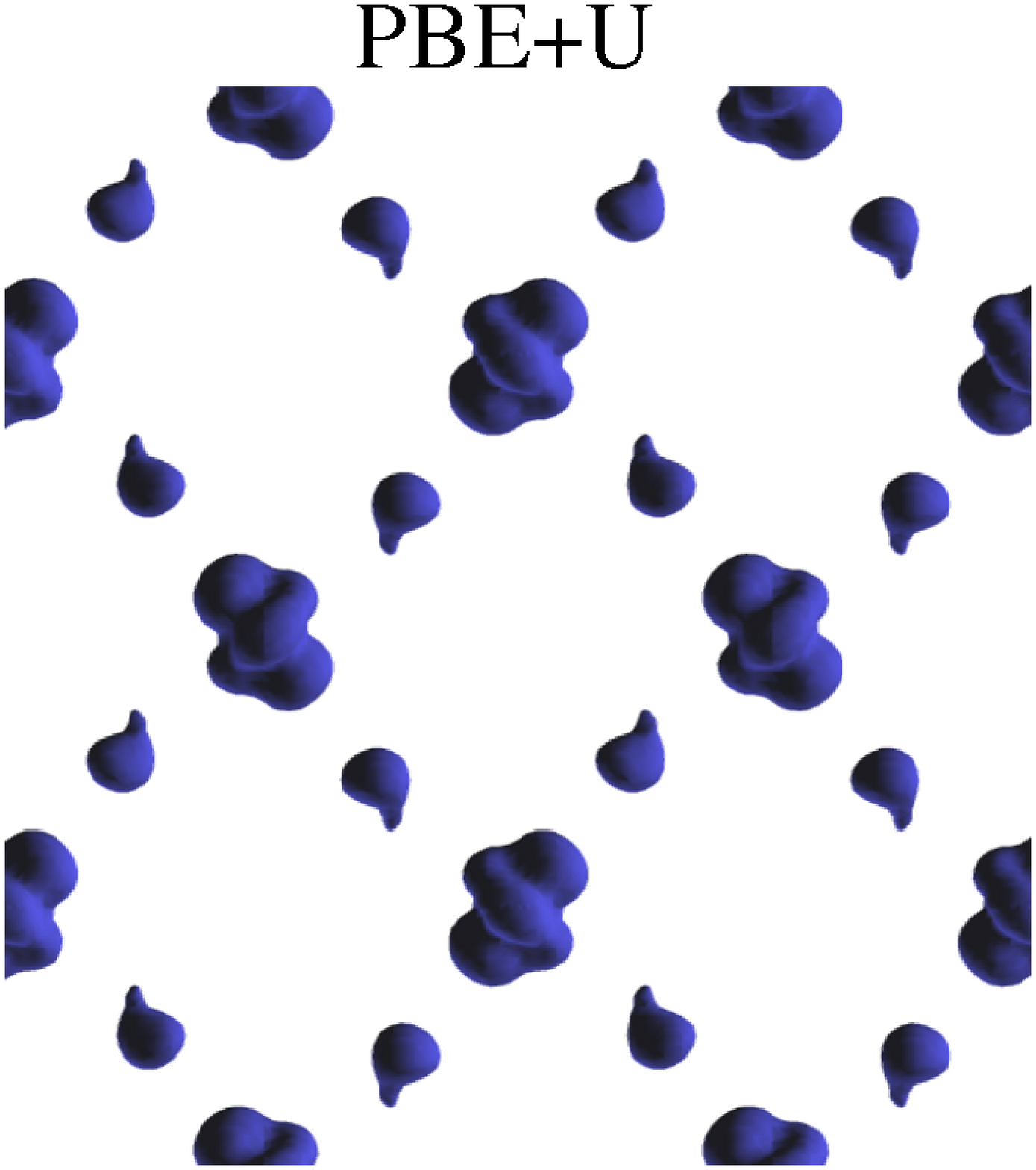}
\hspace{0.1cm}
\includegraphics[clip,width=0.22\textwidth]{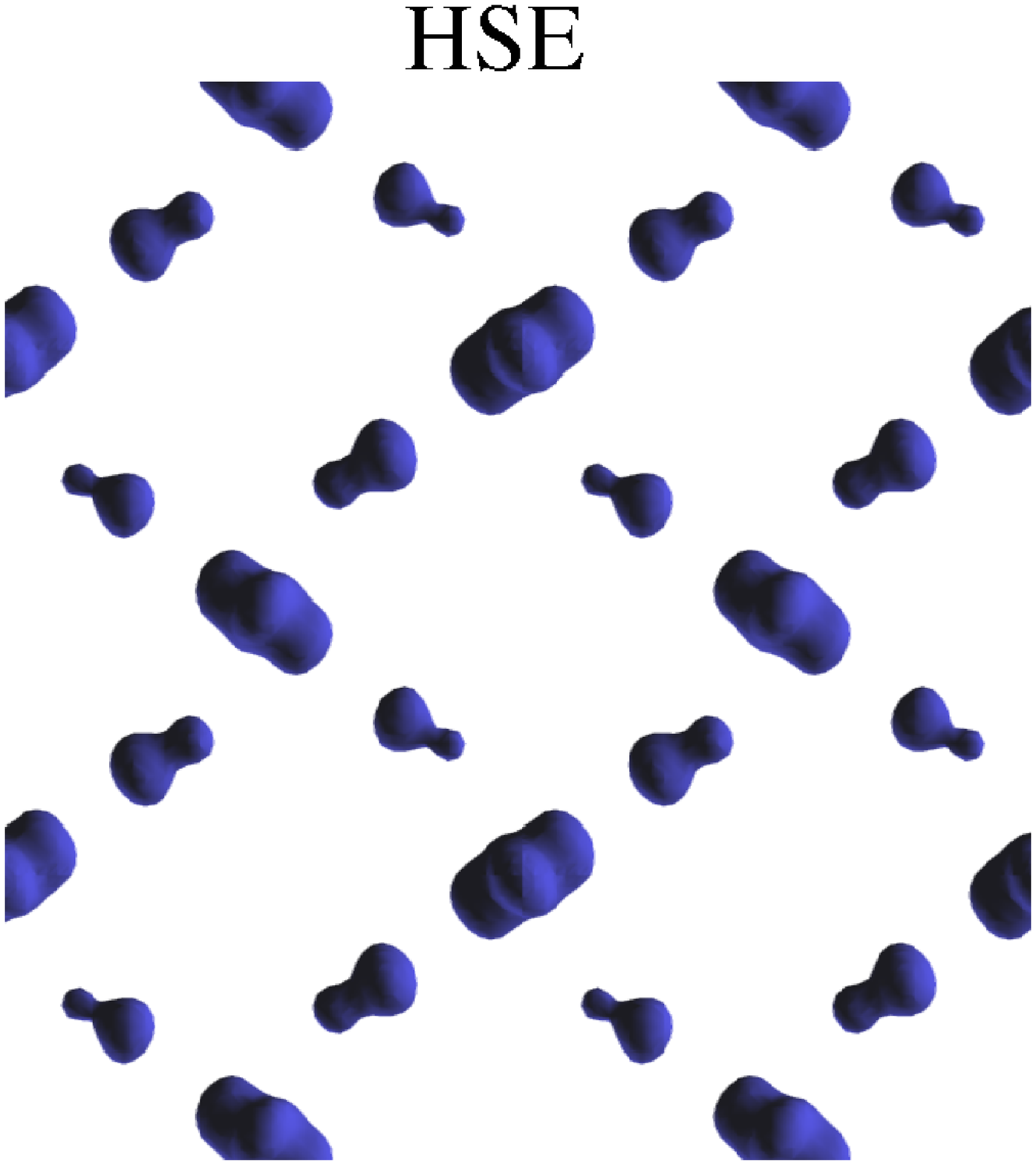}
\hspace{0.1cm}
\includegraphics[clip,width=0.22\textwidth]{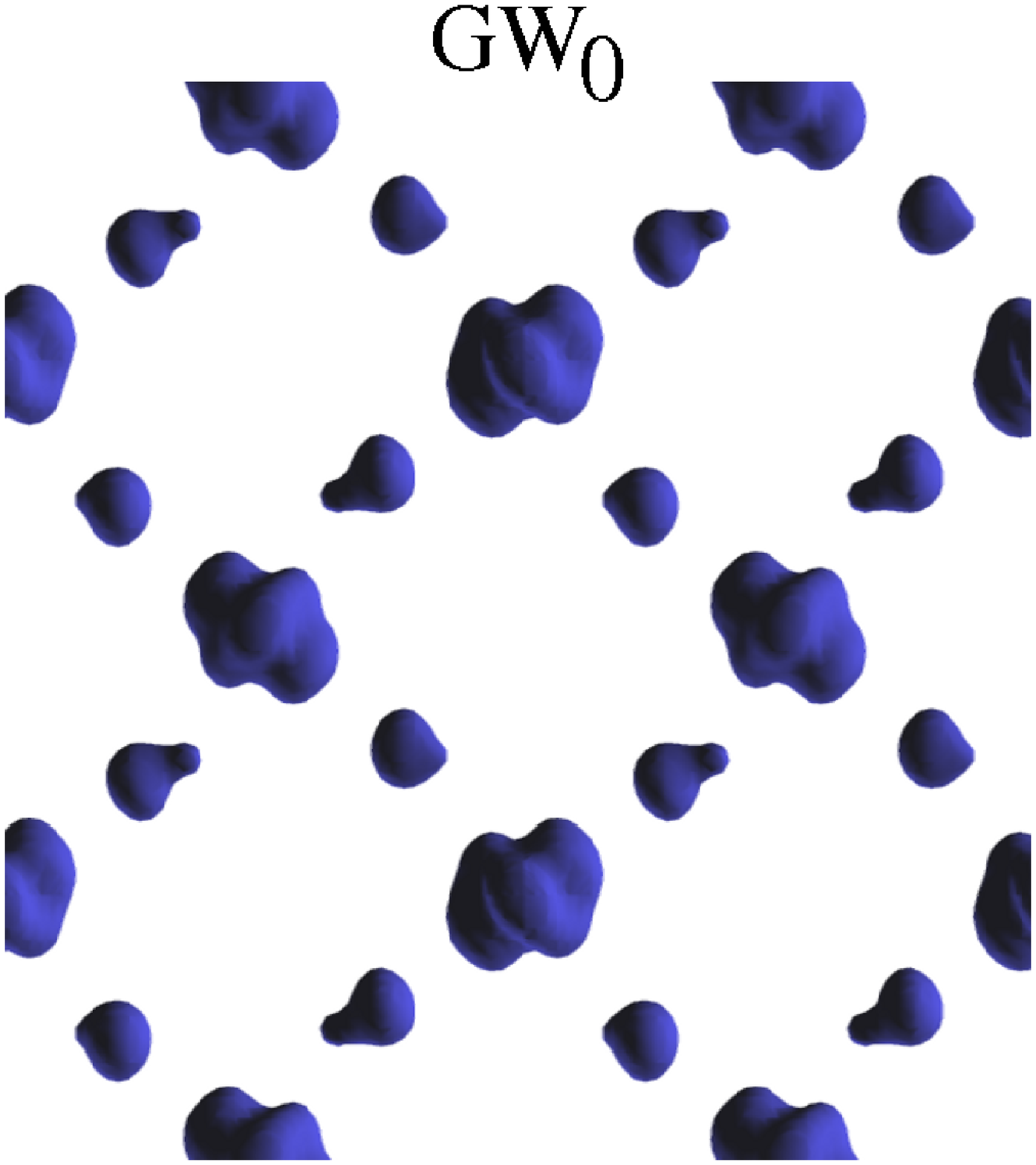}
\caption{Charge density isosurfaces of the highest occupied $e_g$ orbitals (from E$_F$ to the lower 
energy bound as defined in Fig.\ref{fig:bs-mlwf}) showing the orbitally ordered state of LaMnO$_3$ 
obtained using the different methodologies employed in this study.}
  \label{fig:oo}
\end{figure}

To further demonstrate the similarities between MLWFs calculated at
different levels of theory, we show in \fref{fig:rs-mlwf} the real
space representation of the 2 MLWFs localized at a certain Mn site,
projected on the $xy$ plane. The dominant $e_{g}$ character at the
central Mn site together with the ``hybridization tails'' of mostly
$p$ character at the surrounding O sites is clearly visible for all
MLWFs and methods. For the local majority spin MLWFs (1st and 3rd
row), there is essentially no visible difference in orbital character
between PBE and PBE$+U$, only the O($p$) tails are marginally stronger
if the Hubbard $U$ correction is applied. At the HSE level, both local
majority MLWFs exhibit significant $x$/$y$ asymmetry, leading to more
pronounced O($p$) hybridization tails along the short and long Mn-O
bond for the $\vert{3z^2-r^2}\rangle$-like and
$\vert{x^2-y^2}\rangle$-like function, respectively. Within GW$_0$, the
central $e_{g}$-like part as well as the O($p$) tails are less
asymmetric than for HSE, and appear similar to PBE/PBE$+U$ for both
local majority MLWFs. In comparison with the local majority MLWFs, the
O($p$) hybridization tails of the local minority MLWFs (2nd and 4th
row) are generally less pronounced. There is no significant difference
between the local minority spin MLWFs calculated using the different
methods.  Even at the PBE$+U$ and HSE levels, for which the admixture
of the $t_{2g}$ character could not be controlled systematically,
there is no apparent difference in comparison with PBE and GW$_0$.
The orbitally ordered states resulting from this set of MLWFs basis set is shown in
Fig.\ref{fig:OOc-mlwf} in terms of charge density isosurfaces of the highest occupied and
lower unoccupied orbitals associated to the $e_g$ bands below and above $\rm E_F$ in the 
lower energy window as defined in Fig. \ref{fig:bs-mlwf}. This plot clearly show the 
staggered ordering at neighbouring Mn sites and the significant $p-d$ hybridization at 
the oxygen sites.
As a comparison we provide in Fig. \ref{fig:oo} the corresponding staggered ordering associated to 
the highest occupied $e_g$-like bands as obtained from the full {\em ab initio} self-consistent charge 
density (without downfolding) within the various methods employed in the present study. The similarities 
between the {\em ab initio} and wannierized orbital ordering is a further demonstration of the quality and 
reliability of our wannierization procedure.

\subsection{\label{ssec:res-tb} Tight binding model Hamiltonian}


As already mentioned in the introduction, the electronic Hamiltonian
of the $e_{g}$ manifold in manganites is generally described within
the TB formalism as a sum of the kinetic energy
$\hat{H}_{\mathrm{kin}}$ and several local interaction terms, the
Hund's rule coupling to the $t_{2g}$ core spin
$\hat{H}_{\mathrm{Hund}}$, the JT coupling to the oxygen octahedra
distortion $\hat{H}_{\mathrm{JT}}$, and eventually the
electron-electron interaction $\hat{H}_{\mathrm{e-e}}$, which can be
written as (see
e.g. Refs.~\cite{dagotto,ahn00,ederer07,lin08,kovacik10,kovacik11})
\numparts
\begin{eqnarray}
  \hat{H}_{\mathrm{kin}}= -\sum_{a,b,\bi{R},\Delta\bi{R},\sigma}
  \hat{c}_{\sigma,a\left(\bi{R}+\Delta\bi{R}\right)}^{\dagger}
  t_{\sigma,a\left(\bi{R}+\Delta\bi{R}\right)b\left(\bi{R}\right)}
  \hat{c}_{\sigma,b\left(\bi{R}\right)}\,,
  \label{eq:tb-kin}\\
  \hat{H}_{\mathrm{Hund}}=
  -J_{\mathrm{H}}\sum_{\bi{R}}
  \bi{S_R}
  \sum_{a,\sigma,\sigma'}
  \hat{c}_{\sigma,a\left(\bi{R}\right)}^{\dagger}
  \btau_{\sigma\sigma'}
  \hat{c}_{\sigma',a\left(\bi{R}\right)}\,,
  \label{eq:tb-hund}\\
  \hat{H}_{\mathrm{JT}}=
  -\lambda\sum_{a,b,\bi{R},i,\sigma}
  \hat{c}_{\sigma,a\left(\bi{R}\right)}^{\dagger}
  Q^{i}_{\bi{R}}\tau_{ab}^{i}
  \hat{c}_{\sigma,b\left(\bi{R}\right)}\,,
  \label{eq:tb-jt}\\
  \hat{H}_{\mathrm{e-e}}=
  \case{1}{2}\sum_{a,b,c,d,\sigma,\sigma'}
  U_{abcd}
  \hat{c}_{\sigma,a(\bi{R})}^{\dagger}
  \hat{c}_{\sigma',b(\bi{R})}^{\dagger}
  \hat{c}_{\sigma',d(\bi{R})}
  \hat{c}_{\sigma,c(\bi{R})}\,.
  \label{eq:tb-ee}
\end{eqnarray}
\endnumparts Here, $\hat{c}_{\sigma,a\left(\bi{R}\right)}$ and
$\hat{c}_{\sigma,a\left(\bi{R}\right)}^{\dagger}$ are the annihilation
and creation operators associated with orbital $\vert{a}\rangle$ and
spin $\sigma$, centered at site $\bi{R}$. Furthermore,
$t_{\sigma,a\left(\bi{R}+\Delta\bi{R}\right)b\left(\bi{R}\right)}$ are
the hopping amplitudes between orbitals at site $\bi{R}$ and
$\bi{R}+\Delta\bi{R}$, $\tau_{ab}^{i}$ are the standard Pauli
matrices, $J_{\mathrm{H}}$ is the Hund's rule coupling strength,
$\bi{S_R}$ is the normalized $t_{2g}$ core spin at site $\bi{R}$,
$\lambda$ is the JT coupling constant, and $Q^{i}_{\bi{R}}$ is the
amplitude of a particular JT mode $(i=\{x,z\})$. In our TB analysis we
will only consider the electron-electron interaction within a
mean-field approximation and use a simplified version of
Eq.~(\ref{eq:tb-ee}) corresponding to
$U_{aaaa}=U_{abab}=U_{\mathrm{W}}$ and all other interaction matrix
elements set to zero, which is consistent with the PBE+$U$ treatment
according to Dudarev et al.~\cite{dudarev}. The resulting shift in the
one-electron potential due to the electron-electron interaction then
becomes
\begin{equation}\label{eq:tb-hubpot}
  V_{\sigma,ab}=U_{\mathrm{W}}\left(\case{1}{2}\delta_{ab}-n_{\sigma,ab}\right)\,,
\end{equation}
where $U_{\mathrm{W}}$ is the Hubbard parameter in the basis of MLWFs
and $n_{\sigma,ab}$ are the corresponding occupation matrix
elements.\footnote{Here and in the following we often suppress either
  site or spin indeces or both of them, unless the corresponding
  values do not become clear from the context. Apart from the hopping
  amplitudes all quantities are diagonal in site index. In addition,
  for the collinear configurations of core-spins $\bi{S}_\bi{R}$
  considered here, the Hamiltonian and all quantities involved are
  also diagonal with respect to the global spin projection.}

The model parameters ($t_{\sigma,a(\bi{R}+\Delta\bi{R})b(\bi{R})}$,
$J_{\mathrm{H}}$, $\lambda$, $U_{\mathrm{W}}$) which determine the TB
model Hamiltonian can in principle be obtained from the Hamiltonian
matrix elements $h_{nm}^{\Delta\bi{T}}$ in the MLWF basis. We note
that $\Delta\bi{T}$ in \eref{eq:tbh} refers to lattice translations
whereas $\Delta\bi{R}$ in \eref{eq:tb-kin} refers to the relative
position with respect to the lattice of Mn sites. In the following we
will therefore use the following simplified notation:
$h^{\Delta\bi{T}}_{nm} \rightarrow h^{\Delta\bi{T}}_{a\bi{R},b\bi{R}'}
\rightarrow h^{\Delta\bi{R}}_{ab}$ where
$\Delta\bi{R}=\bi{R}'-\bi{R}+\Delta\bi{T}$. Then $a$ and $b$
correspond to the two effective $e_{g}$ orbitals centered at
individual Mn sites separated by $\Delta\bi{R}$. In order to further
simplify the notation for the hopping amplitudes, we choose one Mn
site as the origin ($\bi{R}=\mathbf{0}$) and align the $x$ and $y$
axes of our coordinate system with the directions corresponding to the
long and short Mn-O bond of the JT($Q^x$) mode, respectively. We then
define the vectors $\hat{\bi{x}}$, $\hat{\bi{y}}$, $\hat{\bi{z}}$
according to the nearest-neighbor spacing of the Mn sites along the
respective axes.

Our TB parameterization is based on the procedure described by some of
the authors in~\cite{kovacik10}, with certain modifications, explained
in the following. In~\cite{kovacik10} it was shown that, at least at
the PBE level, the influence of an individual structural distortion
(JT or GFO) on the Hamiltonian matrix elements $h_{ab}^{\Delta\bi{R}}$
expressed in the basis of $e_{g}$-like MLWFs is to a great extent
independent from the other distortion, and that furthermore the
magnetic configuration has only a weak influence on the resulting
model parameters. The TB parameterization was therefore based on
various model structures with both FM (which always leads to a
metallic system) and A-AFM order, with individual structural
distortion modes frozen in. Due to the significantly increased
computational cost of the HSE and GW$_0$ methods in comparison with PBE
(in particular for the metallic state for which a dense k-points mesh
is required to achieve a well converged solution), it is desirable to
derive the TB parameters from as few (and if possible insulating)
model structures as possible. In the present study, we therefore
construct the TB parameterization from only two crystal structures:
the purely JT($Q^x$) distorted structure and the experimental $Pbnm$
structure, in both cases with A-AFM order, which then yields an
insulating solution. As we will show in \tref{tab:tbparI}, the TB
parameters derived in this way at the PBE level deviate only
marginally from the parameters found in the previous
study~\cite{kovacik10}.

In the following we describe the modified method we use to construct
parameters of the TB model \eref{eq:tb-kin}-\eref{eq:tb-ee}. Many of
the simplifications on which our effective TB description of LaMnO$_3$
is based can be understood from the MLWF matrix elements shown in \fref{fig:me-mlwf}
and will be discussed in the remainder of this section. We will first consider an
effectively ``noninteracting'' case in which we neglect the term
$\hat{H}_{\mathrm{e-e}}$ and consider how the more sophisticated
beyond-PBE treatment of the exchange-correlation kernel affects the
hopping, JT, and GFO-related parameters.  We name this approach Model
1. Then, we discuss an alternative way which involves an explicit
treatment of $\hat{H}_{\mathrm{e-e}}$ in the model Hamiltonian within
mean-field approximation. This allows us to obtain estimates for the
corresponding on-site interaction parameters, by keeping the
conventional PBE description as reference. We call this Model
2. Further technical details can be found in the Appendix.


\subsubsection{\label{sssec:ni-tb} TB parameterization with implicit
  el-el interaction: Model 1.}

\begin{figure}
  \includegraphics[clip,width=1.0\textwidth]{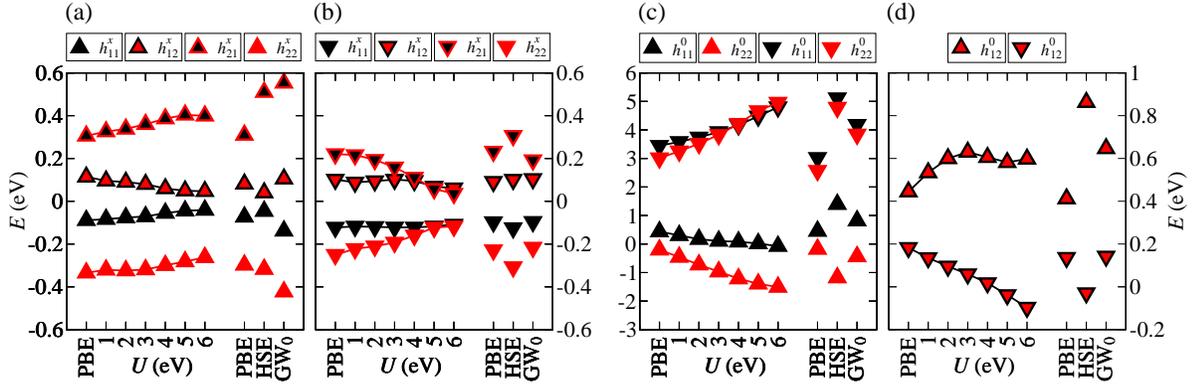}
  \caption{Hamiltonian matrix elements in the basis of MLWFs for the
    experimental $Pbnm$ structure: nearest-neighbor terms
    corresponding to local majority (a) and minority (b) spin
    projection, diagonal (c) and off-diagonal (d) on-site terms. Local
    majority and minority spin projections are indicated by up and
    down triangles, respectively. Left/right parts of the horizontal
    axis corresponds to PWscf/VASP results.}
  \label{fig:me-mlwf}
\end{figure}

As shown in \cite{kovacik10} for the PBE case, good agreement between
an effective $e_g$ TB model and the underlying Kohn-Sham band
structure can be achieved by considering hopping only between nearest
neighbor Mn sites, next-nearest Mn neighbors, and second-nearest Mn
neighbors along the $x$, $y$, and $z$ axes, described by parameters
$t^{ss}$, $t^{xy}$, and $t^{2z}$, respectively (see Appendix). Thereby
it is necessary to take into account the spin dependence of the
nearest neighbor hopping amplitudes. This can also be seen from
\fref{fig:me-mlwf}(a) and (b), where (for PBE) the difference between
$(h_{aa}^{x})^{\uparrow}$ and $(h_{aa}^{x})^{\downarrow}$ (from which
$t^{\uparrow\uparrow}$ and $t^{\downarrow\downarrow}$ are calculated
using \eref{eq:tb-tz}) is indeed significant. On the other hand, the
further neighbor hoppings ($t^{xy}$ and $t^{2z}$) show only negligible
spin dependence, and are therefore calculated from the corresponding
spin averaged Hamiltonian matrix elements. We note that $s$ in
$t^{ss}$ should be read as a \emph{local} spin index (i.e. relative to
the orientation of the local core-spin $\bi{S}_\bi{R}$) corresponding
to the sites between the electron hops, which can have the values $\pm
1$ corresponding to $\uparrow$/$\downarrow$. The parameters
$t^{\uparrow\uparrow}$ and $t^{\downarrow\downarrow}$ thus represent
hopping amplitudes within FM ordered planes. As a result of the GFO
distortion, $t^{\uparrow\uparrow}$ and $t^{\downarrow\downarrow}$ are
reduced by a factor $(1-\eta_{t}^{s})$, where $\eta_{t}^{s}$ is
determined from the ratio of the $t^{ss}$ calculated for the $Pbnm$
and JT($Q^{x}$) structures (see \eref{eq:tb-etat} in Appendix). The
hopping amplitude $t^{\uparrow\downarrow}$ between A-AFM ordered
planes is then calculated as average of $t^{\uparrow\uparrow}$ and
$t^{\downarrow\downarrow}$. As also shown in \cite{kovacik10}, the JT
distortion induces a strong splitting between the nondiagonal elements
of the nearest-neighbor hopping matrix within the $xy$ plane (see the
differences between $h_{12}^{x}$ and $h_{21}^{x}$ in
\fref{fig:me-mlwf}(a,b)), which is parameterized via a non-local JT
coupling strength $\widetilde{\lambda}$ (see \eref{eq:tb-lambdatilde}
in Appendix).

Within model 1 only two contributions to the on-site part of the TB
Hamiltonian are considered: the Hund's rule coupling
$\hat{H}_{\mathrm{Hund}}$ and the Jahn-Teller coupling
$\hat{H}_{\mathrm{JT}}$. The strength of the Hund's rule coupling
$J_{\mathrm{H}}$ is determined from the spin splitting of the on-site
diagonal matrix elements $h^0_{aa}$ for the $Pbnm$ structure, averaged
over both orbitals (see Eq.~\eref{eq:tb-J}). The JT coupling strength
$\lambda^s$ for local spin-projection $s$ is determined according to
Eq.~\eref{eq:tb-jt} from the splitting of the eigenvalues of the
on-site Hamiltonian matrix $\uuline{{h}}^{0}$ and the JT
amplitude $Q^x$ for the purely JT($Q^{x}$) distorted structure. As can
be seen from \fref{fig:me-mlwf}(c,d), the corresponding matrix
elements are strongly spin-dependent, leading to large differences in
the corresponding JT coupling constants. Similar to the hopping
amplitudes, $\lambda^{s}$ is reduced by a factor
$\left({1-\eta_{\lambda}^s}\right)$ due to the GFO distortion, which
is determined from the ratio between $\lambda^s$ calculated for the
$Pbnm$ and JT($Q^{x}$) structures.
%

\Tref{tab:tbparI} lists the obtained TB parameters corresponding to
Model 1 calculated within the various levels of approximation. Both
hopping amplitudes and JT coupling strength correspond to the case
without GFO distortion. It can be seen from the first two rows of
\tref{tab:tbparI} that the parameterization we use in the present
study yields only marginal differences for the PBE hopping parameters
and Hund's rule coupling in comparison with~\cite{kovacik10}. This
corroborates the quality of our TB parameterization based on only two
structures (JT($Q^x$) and $Pbnm$ with A-AFM order). Note, that here we
use a crystal structure derived from low-temperature
measurements~\cite{elemans71}, whereas in \cite{kovacik10} the room
temperature measurements of Ref.~\cite{norby95} have been used. The JT
coupling parameters differ slightly more from \cite{kovacik10}, due to
the revised definition of $\lambda^{s}$ used in the present
study. Another important change arises from the use of 3 separate GFO
reduction factors $\eta_t^{\uparrow}$, $\eta_t^{\downarrow}$, and
$\eta_{\lambda}$, instead of using one averaged value as it was done
in \cite{kovacik10}), which provides a more accurate TB description of
the MLWF bands. It can be also seen from \tref{tab:tbparI} that at the
PBE level, there is essentially no difference between the hopping
amplitudes calculated using PWscf and VASP. There is a 12~\%
difference in $J_{\mathrm{H}}$ between PBE(VASP) and PBE(PWscf), which
could be related to the noticeable differences in the energetics of
the various magnetic configurations discussed earlier.

\begin{table}
  \caption{The TB model parameters as derived from PBE and beyond-PBE
    band structures (Model 1, in PBE+U we used U=3 eV).  Since the PBE$+U$ values of
    $\eta_t^{\downarrow}$ and $\lambda^\downarrow$ are unreliable (see
    text), we use the corresponding PBE values (in brackets) to
    compute the TB bands displayed in \fref{fig:bs-tb}.  Units:
    $t^{\uparrow\uparrow}$, $t^{\downarrow\downarrow}$, $t^{xy}$,
    $t^{2z}$ in meV; $\widetilde{\lambda}$ in meV/\AA;
    $J_{\mathrm{H}}$ in eV; $\lambda^\uparrow$, $\lambda^\downarrow$
    in eV/\AA; $\eta_t^{\uparrow}$, $\eta_t^{\downarrow}$,
    $\eta_{\lambda}$ are unit-less.}  \lineup
  \begin{indented}
  \item[]
  \centering
  \begin{tabular}{@{}lccccccccccc}
\br
& \centre{7}{Hopping parameters} & \centre{4}{On-site parameters}\\\ns\ns
& \crule{7} & \crule{4}\\
&$t^{\uparrow\uparrow}$%
&$t^{\downarrow\downarrow}$%
&$\widetilde{\lambda}$%
&$t^{xy}$%
&$t^{2z}$%
&$\eta_t^{\uparrow}$%
&$\eta_t^{\downarrow}$%
&$J_{\mathrm{H}}$%
&$\lambda^\uparrow$%
&$\lambda^\downarrow$%
&$\eta_{\lambda}$\\
\mr\bs
\multicolumn{12}{c}{PWscf}\\\bs
PBE \cite{kovacik10}  &648&512&530&18&30&0.26&0.26&1.50&\03.19&1.33&0.26\\
PBE                   &632&512&523&12&51&0.28&0.39&1.56&\03.35&1.07&0.22\\
PBE+$U$               &748&482&516&12&51&0.41&(0.39)&2.16&\05.22&(1.07)&0.21\\\bs
\multicolumn{12}{c}{VASP}\\\bs
PBE                   &630&503&516&13&50&0.35&0.42&1.33&\03.21&1.02&0.23\\
HSE                   &750&497&707&13&50&0.40&0.20&2.42& 10.25&0.96&0.28\\
GW$_0$                &746&469&490&13&50&0.24&0.41&1.90&\04.43&0.88&0.04\\
\br
  \end{tabular}
  \end{indented}
  \label{tab:tbparI}
\end{table}

\begin{figure}
  \centering
  \includegraphics[clip,width=\textwidth]{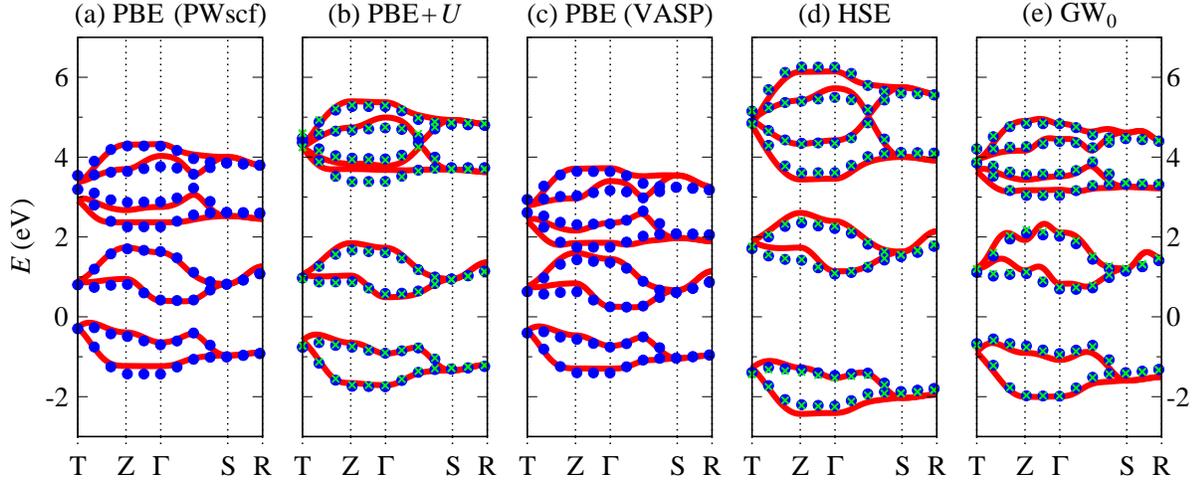}
  \caption{Comparison of the band dispersion corresponding to MLWFs
    (red lines), the TB Model 1 using parameters given in
    \tref{tab:tbparI} (blue circles), and the TB Model 2 with
    interaction parameters given in \tref{tab:tbparII} (green
    crosses).}
  \label{fig:bs-tb}
\end{figure}

Comparing the parameters obtained from the beyond-PBE methods with the
pure PBE case, we observe that the hopping parameter
$t^{\uparrow\uparrow}$ is generally increased in all beyond-PBE
methods. As was shown in~\cite{kovacik11}, this can be understood
within an extended nearest neighbor TB model including both Mn($d$)
and O($p$) states, from which an effective $e_g$-only model can be
derived in the limit of large energy separation $\varepsilon_{dp}$
between the $d$ and $p$ orbitals. The effective hopping
$t_{dd}^{\mathrm{eff}}$ in the $e_g$ model is then given in terms of
the nearest neighbor hopping amplitude $t_{dp}$ of the extended
$d$-$p$ model as
$t_{dd}^{\mathrm{eff}}=t_{dp}^{2}/\varepsilon_{dp}$. The increase of
$t^{\uparrow\uparrow}$ is therefore consistent with the observation
that all beyond-PBE methods lower the $e_g$ bands relative to the
lower-lying oxygen $p$ bands. 
The small decrease of $t^{\downarrow\downarrow}$ within PBE$+U$ (for small
values of $U \lesssim 2$~eV) can be explained in the same way, since
here the corresponding energy separation between O($p$) and Mn($e_g$)
increases. The JT parameter $\widetilde{\lambda}$ is generally very
similar for PBE, PBE$+U$, and GW$_0$, while a strong enhancement of
$\widetilde{\lambda}$ can be seen for HSE, which is consistent with
the strong $x/y$ asymmetry of the corresponding MLWFs seen in
\fref{fig:rs-mlwf}(c). Since the changes of the already rather small
further-neighbor hoppings within the beyond-PBE methods are very
small, we use the corresponding PBE values for simplicity. The GFO
reduction factors for the hopping amplitudes, $\eta_{t}^{\uparrow}$
and $\eta_{t}^{\downarrow}$, are slightly decreased within GW$_0$, whereas
$\eta_{t}^{\uparrow}$ is increased for PBE$+U$ and HSE, and
$\eta_{t}^{\downarrow}$ is strongly decreased in HSE. Due to the
strong mixing between minority spin $e_{g}$ and $t_{2g}$ bands within
PBE+$U$, which was already discussed in section~\ref{ssec:mlwf} (see
also \fref{fig:bs-tb}(b)), the determination of $\eta_t^\downarrow$ is
rather unreliable in this case, and we therefore use the corresponding
PBE value. We note that the same effect also leads to the strong
changes in the l cal minority hopping matri xelements within the $xy$
plane calculated within PBE$+U$ for $U \gtrsim 3$~eV (see
\fref{fig:me-mlwf}(b)). Using the HSE and GW$_0$ methods we do not
encounter this problem.

For all beyond-PBE methods, a significant increase of $J_{\mathrm{H}}$
and $\lambda^{\uparrow}$ can be observed, which in the TB model gives
rise to an increase of the spin splitting and the band gap,
respectively. The change of $\lambda^{\downarrow}$ compared to PBE is
very small for both HSE and GW$_0$. Due to the inaccurate treatment of the
minority spin bands, PBE$+U$ gives an unrealistically small value of
$\lambda^{\downarrow}=0.30$~eV/\AA, which we therefore substitute with
the corresponding PBE value. While $\eta_{\lambda}$ does not change
significantly for small values of the Hubbard $U$, a small increase
(significant decrease) is observed for HSE (GW$_0$).

To assess the quality of our parameterization we now use the TB
parameters tabulated in \tref{tab:tbparI} to compute the resulting
$e_g$ band structure.  In \fref{fig:bs-tb}(a) and (c), we compare the
band dispersions of the TB model (blue filled circles) and the MLWFs
(thick red lines) for the experimental $Pbnm$ structure within the PBE
approximation. Despite the many simplifications made in the
construction of the model parameters, the TB model can reproduce the
MLWF bands to a remarkable accuracy (for both PWscf and VASP). The
reliability of the beyond-PBE TB representation can be appreciated by
the overall excellent match between the TB and MLWFs bands shown in
\fref{fig:bs-tb}(b), (d) and (e), which exhibit the same quality as
observed at the PBE level. This is particularly true for the band gap,
whose method-dependent changes (see \tref{tab:1}) are perfectly
reflected in the TB description.\footnote{The MLWF and TB bands were
  aligned by minimizing a mean deviation which was calculated as an
  average of the corresponding eigenvalue differences over all bands
  and k-points. The maximum and mean deviation is very similar for all
  methods and does not exceed 0.37 and 0.12~eV, respectively.}

\subsubsection{\label{sssec:i-tb} TB parameterization with explicit
  el-el interaction: Model 2.}

Now, we turn our attention on the alternative TB parameterization in
which we attempt to treat the modifications induced by the beyond-PBE
methods as a perturbation to the ``noninteracting'' PBE description by
explicitly considering the el-el interaction \eref{eq:tb-ee} and using
the simplified mean-field approximation \eref{eq:tb-hubpot} in the TB
Hamiltonian. It is clear from the discussion in the preceeding section
that it is not straightforward to parameterize the hopping amplitudes
in terms of $U_{\mathrm{W}}$. We will therefore limit ourselves to
analyzing the effect of \eref{eq:tb-hubpot} on the local Hamiltonian,
which can be represented as 2$\times$2 matrix in terms of the two
local $e_g$ states in the following form:
\begin{equation}
  \label{eq:tb-local}
  \uuline{{H}}{}_{\,\mathrm{local}}^s = \uuline{\tilde{{H}}}{}_{\,0}^s - U_{\mathrm{W}} \uuline{{n}}^s
\end{equation}
with
\begin{equation}
  \uuline{\tilde{{H}}}{}_{\,0}^s = \uuline{{1}} \left( \case{1}{2} U_{\mathrm{W}} - J_H
    \cdot s \right) - \lambda^s Q^x \uuline{\tau}{}^x - \lambda^s Q^z \uuline{\tau}^z \,.
\end{equation}
By identifying \eref{eq:tb-local} with the corresponding MLWF matrix,
we obtain the local spin splitting as a combination of Hund's rule
coupling and el-el interaction:
\begin{equation}
  \left(h_{aa}^{0}\right)^{\downarrow}-\left(h_{aa}^{0}\right)^{\uparrow}=
  U_{\mathrm{W}}^{(J)}\big(n_{aa}^{\uparrow}-n_{aa}^{\downarrow}\big)
  +2J_{\mathrm{H}}^{(\mathrm{PBE})}\,,
  \label{eq:tb-UWJder}
\end{equation}
from which we can calculate $U_{\mathrm{W}}^{(J)}$ by averaging over
the two orbital characters and using the previously determined PBE
value for the Hund's rule coupling. Thereby, the occupation matrix
elements in the basis of MLWFs are calculated as
\begin{equation}\label{eq:n-mlwf}
  n_{nm} = \int_{-\infty}^{E_{\mathrm{F}}} \mathrm{d}\epsilon \int_\mathrm{BZ} \mathrm{d}\bi{k}
  \sum_{l} \left(U^{(\bi{k})}_{lm}\right)^* \delta(\epsilon-\epsilon_{l\bi{k}}) \, U^{(\bi{k})}_{ln}
  \,,
\end{equation}
where $E_{\mathrm{F}}$ is the Fermi energy.

In a similar way we can obtain another estimate for the Hubbard
parameter, $U_{\mathrm{W}}^{(\lambda)}$, from the total JT induced
splitting within the majority spin $e_g$ orbital manifold, expressed
through the difference in eigenvalues of the local Hamiltonian:
\begin{equation}
  \Delta{\varepsilon}^{\uparrow} = 2 \lambda^{\uparrow(\mathrm{PBE})}
  \sqrt{(Q^x)^2+(Q^z)^2} + U_{\mathrm{W}}^{(\lambda)}\Delta{n}^{\uparrow} =
  \Delta{\varepsilon}^{\uparrow(\mathrm{PBE})} +
  U_{\mathrm{W}}^{(\lambda)}\Delta{n}^{\uparrow} \,.
  \label{eq:tb-UWlambdader}
\end{equation}
Here, $\Delta{n}^{\uparrow}$ is the difference in majority spin
eigenvalues of the MLWF occupation matrix and we have used the
observation that, to a very good approximation, both
$\uuline{\tilde{{H}}}{}_{\,0}$ and $\uuline{{n}}$ can be diagonalized
by the same unitary transformation.  The difference between the
corresponding transformation angles is less than $0.6^{\circ}$ for the
$Pbnm$ structure. Since the difference is somewhat larger for the
JT($Q^x$) structure (up to $\approx 6^{\circ}$) we derive the
interaction parameter $U_{\mathrm{W}}^{(\lambda)}$ from the MLWF
Hamiltonian of the $Pbnm$ structure. The resulting values of
$U_{\mathrm{W}}^{(J)}$ and $U_{\mathrm{W}}^{(\lambda)}$ are given in
\tref{tab:tbparII}.
\begin{table}
  \caption{\label{tab:tbparII}The interaction parameters determined in
    Model 2.  Note, that in Model 2 the on-site parameters are set to
    the PBE values while the hopping parameters are set to the values
    given in \tref{tab:tbparI}.  Units: all quantities are in eV
    except $\Delta{n}^{\uparrow}$ which is unit-less.}
  \begin{indented}
  \item[]
    \begin{tabular}{lcccccc}
      \br
      &&&& \centre{3}{Interaction parameters}\\\ns\ns
      &&&& \crule{3}\\
      &$J_{\mathrm{H}}$%
      &$\Delta{\varepsilon}^{\uparrow}$%
      &$\Delta{n}^{\uparrow}$%
      &$U_{\mathrm{W}}^{(J)}$%
      &$U_{\mathrm{W}}^{(\lambda)}$%
      &$\Delta{J_{\mathrm{W}}^{(\lambda)}}$\\
      \mr\bs
      \multicolumn{7}{c}{PWscf}\\\bs
      PBE      &1.56&1.09&0.71& -  & -  & -  \\
      PBE$+U$  &2.16&1.66&0.80&2.40&0.70&0.42\\\bs
      \multicolumn{7}{c}{VASP}\\\bs
      PBE      &1.33&1.04&0.70& -  & -  & -  \\
      HSE      &2.42&3.10&0.89&4.37&2.31&0.51\\
      GW$_0$   &1.90&1.80&0.70&2.30&1.09&0.30\\
      \br
    \end{tabular}
  \end{indented}
\end{table}

It can be seen that within PBE$+U$, the parameter
$U_{\mathrm{W}}^{(J)}$ is almost as large as the value of $U=3$~eV
used for the Hubbard parameter within the PBE$+U$ calculation, whereas
the parameter $U_{\mathrm{W}}^{(\lambda)}$ is significantly smaller
than that. We note that, as discussed in \cite{kovacik11}, the Hubbard
correction within PBE+$U$ is applied to rather localized atomic-like
orbitals, whereas the parameter $U_{\mathrm{W}}$ corresponds to more
extended $e_g$-like Wannier orbitals. The JT splitting is strongly
affected by hybridization with the surrounding oxygen ligands and is
thus quite different for atomic-like and extended Wannier states
\cite{kovacik11}. As a result, $U_{\mathrm{W}}^{(\lambda)}$ is quite
different from the $U$ value used within PBE+$U$, and the smaller
value of $U_{\mathrm{W}}^{(\lambda)}$ can thus be related to the fact
that the electron-electron interaction is more screened in the more
extended effective $e_g$ Wannier orbitals. On the other hand, the
similarity between $U_{\mathrm{W}}^{(J)}$ and the $U$ value used
within PBE+$U$ indicates that the local spin-splitting is more or less
the same for both sets of orbitals, which is consistent with the view
that this splitting is essentially an atomic property. A similar
difference between $U_{\mathrm{W}}^{(J)}$ and
$U_{\mathrm{W}}^{(\lambda)}$ is also observed for HSE and GW$_0$. The
large values of $U_{\mathrm{W}}$ delivered by HSE reflects the larger
spin splitting and band gap in the corresponding band structure
compared to PBE+$U$ and GW$_0$.

The large difference between the two parameters $U_{\mathrm{W}}^{(J)}$
and $U_{\mathrm{W}}^{(\lambda)}$ also indicates that it is not
possible to map the electron-electron interaction  effects manifested in the
on-site matrix corresponding to effective $e_g$ orbitals to only one
interaction parameter while using PBE as ``noninteracting'' reference.
Similar conclusions have already been reached in \cite{kovacik11} for
the PBE+$U$ case. From the current study we can conclude that the
modification of the local spin splitting (described by
$U_{\mathrm{W}}^{(J)}$) and the enhancement of the JT induced orbital
splitting (described by $U_{\mathrm{W}}^{(\lambda)}$) that arise in
the Kohn-Sham or GW$_0$ quasiparticle band structures due to the 
beyond-PBE treatment of exchange and correlation, are not compatible
with a simple mean-field Hubbard-like correction to an otherwise
``non-interacting'' TB Hamiltonian with two effective $e_g$ orbitals
per Mn site and only one parameter describing the electron-electron
interaction. This leads to an important conclusion of the present
study with regard to methods such as LDA+$U$ or LDA+DMFT, which
supplement a ``non-interacting'' Kohn-Sham Hamiltonian with a Hubbard
interaction between a strongly interacting subset of orbitals: using
different methods for obtaining the noninteracting reference can lead
to significant differences, and it is by no means clear whether PBE
(GGA) or even LDA always provides the best starting point for a more
sophisticated treatment of correlation effects. Our results also
emphasize the importance of finding improved ways to account for the
double counting correction when using different electronic structures
as noninteracting reference.

In order to see how, within the limitations discussed in the
preceeding paragraph, a TB Hamiltonian of the form
\eref{eq:tb-kin}-\eref{eq:tb-ee} can reproduce the MLWF band
dispersion, we consider a modified parameterization using
$U_{\mathrm{W}}^{(\lambda)}$ to model the el-el interactions. Since in
that way the correlation-induced increase of the spin splitting is
only partially covered by the el-el term \eref{eq:tb-hubpot}, we
correct this by introducing an ``empirical'' correction to the Hund's
rule coupling:
\begin{equation}\label{eq:tb-DeltaJ}
  \Delta{J_{\mathrm{W}}^{(\lambda)}}=
  J_{\mathrm{H}}-J_{\mathrm{H}}^{(\mathrm{PBE})}-\case{1}{4}{U_{\mathrm{W}}^{(\lambda)}}\,.
\end{equation}
Note, that analogously we could choose $U_{\mathrm{W}}^{(J)}$ as the
el-el interaction parameter and define an appropriate correction to
$\lambda^{\uparrow}$. However, since the fundamental band gap in
LaMnO$_3$ is largely controlled by the JT induced splitting between
occupied and unoccupied $e_g$ bands, and since in a TB model for
LaMnO$_{3}$ it seems most desirable to describe the band gap
correctly, we choose $U_{\mathrm{W}}^{(\lambda)}$ to model the el-el
interactions. If the correction $\Delta{J_{\mathrm{W}}^{(\lambda)}}$
is neglected, the local majority spin bands around the band gap are
still described quite well, even though the splitting with respect to
the local minority spin bands will be underestimated, which might be
acceptable for certain applications.

\Fref{fig:bs-tb} also shows the dispersion calculated from such a
modified TB model with explicit el-el interaction, where the
correlation induced change of the spin splitting and band gap is
described by two interaction parameters, $U_{\mathrm{W}}^{(\lambda)}$
and $\Delta{J_{\mathrm{W}}^{(\lambda)}}$, while $J_{\mathrm{H}}$,
$\lambda^{\uparrow}$, $\lambda^{\downarrow}$, and $\eta_{\lambda}$ are
fixed at their respective PBE values. In addition, the hopping
amplitudes are set to the values given in \tref{tab:tbparI}. The band
dispersions using these sets of parameters (shown as green crosses in
\fref{fig:bs-tb}) again almost perfectly follow the MLWF bands. The
agreement between the bands calculated within the two
parameterizations (Model 1 and 2) also reflects the transferability of
the on-site parameters between the structures with and without the GFO
distortion.

\section{\label{sec:sum} Summary}

In this paper we have presented a general scheme to parameterize,
within a TB picture, the band structure of the prototypical JT
distorted $e_g$ perovskite LaMnO$_3$ by means of a suitable
downfolding of the {\em ab initio} electron dispersion relations onto
a small set of MLWFs. The tabulated TB parameters should provide an
interpretative direction for more sophisticated many-body model
Hamiltonian investigations of similar
systems~\cite{pavarini10,kotliar06,kumar06,yamasaki06}.

By comparing the PBE and beyond-PBE findings we can draw the following
conclusions:

\begin{enumerate}[label=(\roman{*}), ref=(\roman{*})]

\item {\em Ab initio} electronic structure results. 
  We find that all methods consistently find a Mott-Hubbard insulating state.    
  GW$_0$ provides the best agreement with experiments in terms of bandgap value, 
  and both GW$_0$ and HSE convey a satisfactory description of valence and 
  conduction band spectra.
  While in the PBE+$U$ and HSE cases a suitable adjustment of the parameters $U$ and
  $a_{\mathrm{mix}}$ can selectively improve the performance with
  respect to either bandgap or magnetic exchange
  interactions, a universal value that provides all quantities with
  good accuracy cannot be found.  Even though the standard value
  $a_{\mathrm{mix}}=0.25$ in HSE seems to provide rather accurate
  magnetic coupling constants, clearly a smaller $a_{\mathrm{mix}}$ is
  necessary to obtain a better Mott-Hubbard gap. While the two
  different codes used in the present study lead only to marginal
  differences in the Kohn-Sham band structure and the corresponding TB
  parameterization, the relative energies of different magnetic
  configurations depend on subtle details of the used methods, which
  hampers a concise comparison between the different energy functional
  (it should be noted however, that the PAW approach is usually
  considered superior to pure pseudopotential schemes).
  Within VASP a value for the Hubbard $U$ between 2-3~eV leads to
  similar magnetic coupling along $c$ as HSE, but somewhat
  stronger FM coupling within the $ab$ planes. Despite all its
  well-known limitations when applied to strongly-correlated
  materials, PBE does not seem to perform too badly (of course the
  fact the we have used the experimental structure helps in that
  respect, since PBE is known to fail in properly reproducing the JT
  distortion in LaMnO$_3$~\cite{hashimoto}).

\item MLWFs. Despite the difficulties to fully disentangle the
  effective $e_g$ bands from other bands with similar energies, which
  are most pronounced within PBE+$U$ and HSE, the resulting MLWFs and associated 
  ordering (Fig.\ref{fig:oo}) look rather similar 
  and are in good agreement with the precedent plots of Yin\cite{yin06}. 
  This represents a further proof of the quality and reliability of the wannier construction of the 
  $e_g$ $|3z^2-r^2\rangle$ and $|x^2 -y^2\rangle$  orbitals.
  Despite these similarities, the differences in the
  underlying band structures lead to distinct differences in the
  Hamilt nian matri xelements in reciprocal space, and allow for an
  accurate quantitative analysis of the differences between the various
  approximations for the exchange-correlation kernel.

\item TB parameterization. We have demonstrated that the
  methods-derived changes in the TB parameters due to the different
  treatment of the el-el exchange-correlation kernel in conventional
  and beyond-PBE approaches can be accounted for using two different
  routes:
  (a) Model 1
  ($\hat{H}_{\mathrm{TB}}=\hat{H}_{\mathrm{kin}} +
  \hat{H}_{\mathrm{Hund}} + \hat{H}_{\mathrm{JT}}$).  In this model
  the TB Hamiltonian does not explicitly incorporate an el-el
  interaction term.  All changes in the beyond-PBE band structure with
  respect to the ``noninteracting'' PBE bands are integrated in the
  hopping, JT and Hund parameters (in particular
  $t^{\uparrow\uparrow}$, $\lambda^\uparrow$, and $J_{\mathrm{H}}$).
  (b) Model 2
  ($\hat{H}_{\mathrm{TB}}=\hat{H}_{\mathrm{kin}} +
  \hat{H}_{\mathrm{Hund}} + \hat{H}_{\mathrm{JT}} +
  \hat{H}_{\mathrm{e-e}}$).  In this second type of parameterization
  we have build in an el-el term in the TB Hamiltonian explicitly. The
  el-el interaction effects are treated by parameterizing the on site
  Hund and JT parameters into a noninteracting (PBE) and interacting
  (dependent on $U_{\mathrm{W}}^{(\lambda)}$ and
  $U_{\mathrm{W}}^{(J)}$) part. Since we found that
  $U_{\mathrm{W}}^{(\lambda)}$ $\neq$ $U_{\mathrm{W}}^{(J)}$, in order
  to achieve a correct parameterization it is necessary to fix one
  $U_{\mathrm{W}}$ channel ($U_{\mathrm{W}}^{(\lambda)}$) and evaluate
  the changes on the remaining one
  ($\Delta{J_{\mathrm{W}}^{(\lambda)}}$).  Both, Model 1 and 2, yield
  excellent TB bands, essentially overlapping with the underlying
  MLWFs ones.

  We note that the different levels of approximation for the
  non-interacting band structure can lead to significant changes in
  the hopping amplitudes, which cannot easily be accounted for by a
  local double-counting correction. In addition, we have also shown
  that the influence of the beyond-PBE treatment on the model
  parameters of the local Hamiltonian cannot be captured by a simple
  mean-field Hubbard term with only one interaction parameter. For an
  accurate many-body or effective model treatment of LaMnO$_3$ and
  similar materials it thus seems most desirable to start from the
  most realistic single particle band-structure (i.e. not necessarily
  LDA or GGA) and use an appropriate double counting correction. 
  The exact form of such a correction term, however, is still unclear at
  this point.
  A possible alternative solution to correctly treat correlation effects 
  without being contaminated by the double-counting problem is the 
  GW+DMFT scheme, which has recently attracted several research groups and
  will be most likely available in the next future\cite{Biermann03,Karlsson05}.

\end{enumerate}

In summary, we have shown that MLWFs can be constructed efficiently
not only at conventional DFT level (PBE), but also from hybrid
functional (HSE) and quasiparticle (GW$_0$) wavefunctions, through the
creation of an appropriate interface between the electronic structure
code VASP~\cite{vasp1,vasp2} and the publicly available Wannier90
code~\cite{wannier90}. Thereby, we have used the well-established
PW2WANNIER90 interface as benchmark at the PBE and PBE+$U$
level~\cite{espresso}. Given the booming application of hybrid DFT and
GW$_0$ calculations for a wide variety of materials for which the
possibility to describe the relevant physics using a minimal basis set
is important (these include, e. g., Fe-based
superconductors~\cite{wojdel}, cuprates~\cite{rivero10} and
multiferroics~\cite{stroppa10,stroppa11}), the VASP2WANNIER90
interface which allows to construct MLWFs directly from the
widely-used VASP code, will provide a valuable tool for future
research.
From the practical point of view, we have demonstrated that 
MLWFs can be efficiently used to accurately interpolate the HSE and GW$_0$ band structure 
from the coarse uniform k-points mesh to the desirable (and dense) symmetry lines,
thereby remedying the fundamental practical limitation of HSE and GW$_0$ scheme 
in computing energy eigenvalues for selected k-points\cite{2001_souza,hamann09}.
We expect that our study will serve as a reference for
future studies involving MLWFs-based downfolding procedure.

\ack
The authors would like to thank Silvia Picozzi and Alessandro Stroppa
(CNR-SPIN, L'Aquila) for initiating the implementation of the
VASP2WANNIER90 interface and for hosting Martijn Marsman in L'Aquila
where large parts of this interface were written. This work has been
supported by the 7$^{th}$ Framework Programme of the European
Community, within the project ATHENA, by Science Foundation
Ireland under Ref.~SFI-07/YI2/I1051, and by the Austrian FWF within 
the SFB ViCoM (F41). PWscf and VASP calculations have
been performed at the Trinity Centre for High-Performance Computing
(TCHPC) and the Vienna Scientific Cluster (VSC), respectively.

\newpage

\appendix
\section*{\label{app:tbparI}Appendix: Tight binding model parameterization}
\setcounter{section}{1}

In this section, we supply exact definitions for all parameters
included in our TB models and describe how they are obtained from the
MLWF Hamiltonian matrix.

In the following, we express the hopping amplitudes
$t_{\sigma,a\left(\bi{R}+\Delta\bi{R}\right)b\left(\bi{R}\right)}$ as
${2}\times{2}$ matrices with respect to the orbital indices:
$t_{\sigma,a\left(\bi{R}+\Delta\bi{R}\right)b\left(\bi{R}\right)}
\rightarrow \uuline{t}^{ss'}\left(\Delta\bi{R}\right)$. There are 4
different hoppings parameters which we consider here, the
spin-dependent nearest-neighbor hopping $t^{ss}$, the second-nearest
neighbor hopping $t^{xy}$ and the second-nearest neighbor hopping
along the $x$, $y$, $z$ axes $t^{2z}$. The nearest-neighbor hopping
matrix is expressed as \numparts
\begin{eqnarray}
  \uuline{{t}}^{ss'}(\pm\hat{\bi{z}})=
  -\case{1}{2}t^{ss'}
  (\uuline{{1}}+\uuline{\tau}^{z})
  \,,\label{eq:tb-hopz}\\
  \uuline{{t}}^{ss}(\pm\hat{\bi{x}})=
  -\case{1}{4}t^{ss}
  (2\cdot\uuline{{1}}-\sqrt{3}\cdot\uuline{\tau}^{x}-\uuline{\tau}^{z})
  \,,\label{eq:tb-hopx}
\end{eqnarray}
\endnumparts (and analogously for
$\uuline{{t}}^{ss'}(\pm\hat{\bi{y}})$, see~\cite{kovacik10}). The
$t^{ss}$ which determines hopping amplitudes within FM ordered planes
is via \eref{eq:tb-hopx} (from the Hamiltonian matrix elements for the
JT($Q^{x}$) distorted structure) given as
\begin{equation}\label{eq:tb-tz}
  t^{ss}=
  \left(\case{1}{2}h_{11}^x-\case{3}{2}h_{22}^x\right)^{s}\,.
\end{equation}
In the GFO distorted structure, the nearest-neighbor hopping
amplitudes are reduced by a factor $(1-\eta_{t}^{s})$, where the
coefficient $\eta_{t}^{s}$ is calculated as
\begin{equation}\label{eq:tb-etat}
  \eta_{t}^{s}=1-
  \frac{t^{ss}[Pbnm]}{t^{ss}[\mathrm{JT}(Q^x)]}\,.
\end{equation}
The hopping parameters are denoted by the corresponding crystal
structure (in square brackets), for which they are calculated. The
$t^{\uparrow\downarrow}$ parameter, which determines the hopping
amplitude between A-AFM ordered planes, is taken as an average of
$t^{\uparrow\uparrow}$ and $t^{\downarrow\downarrow}$. The JT-induced
splitting of the nondiagonal elements of the hopping matrix within the
$xy$ plane is incorporated as an additional contribution to the
in-plane hopping
\begin{equation}\label{eq:tb-hopjt}
  \Delta\uuline{{t}}^{ss}(\pm\hat{\bi{x}})=
  \widetilde{\lambda}Q^x_{\bi{R}}
  (\mathrm{i}\cdot\uuline{\tau}^{y})
  \,,  
\end{equation}
(and analogously for $\Delta\uuline{{t}}(\pm\hat{\bi{y}})$).
The $\widetilde{\lambda}$ parameter is determined for the JT($Q^x$)
distorted structure as
\begin{equation}\label{eq:tb-lambdatilde}
  \widetilde{\lambda}=
  \frac{1}{2Q^{x}}
  \left(
    \case{1}{2}{\left(h_{12}^{x}-h_{21}^{x}\right)^{\uparrow}}
    +\case{1}{2}{\left(h_{12}^{x}-h_{21}^{x}\right)^{\downarrow}}
  \right)\,,
\end{equation}
and is also reduced by the $(1-\eta_{t}^{s})$ factor in the GFO
distorted structure.
Both, the second nearest-neighbor hopping $t^{xy}$ and the
second-nearest neighbor hopping along the $x$, $y$, $z$ axes are
determined (from the JT($Q^{x}$) distorted structure) as \numparts
\begin{eqnarray}
  t^{xy}=
  -\case{1}{2}\left[(h_{11}^{xy})^{\uparrow}+(h_{11}^{xy})^{\downarrow}\right]\,,\label{eq:tb-txy}\\
  t^{2z}=
  -\case{1}{2}\left[(h_{11}^{2z})^{\uparrow}+(h_{11}^{2z})^{\downarrow}\right]\,.\label{eq:tb-t2z}
\end{eqnarray}
\endnumparts
The matrices related to the $t^{xy}$ hopping parameter are then given
by (see e.g. \cite{ederer07})\numparts
\begin{eqnarray}
  \uuline{{t}}(\pm\hat{\bi{x}}\pm\hat{\bi{z}})=
  -t^{xy}
  (-\uuline{{1}}+\sqrt{3}\cdot\uuline{\tau}^{x}-\uuline{\tau}^{z})
  \,,\label{eq:tb-hopxz}\\
  \uuline{{t}}(\pm\hat{\bi{x}}\pm\hat{\bi{y}})=
  -t^{xy}
  (-\uuline{{1}}+2\cdot\uuline{\tau}^{z})
  \,,\label{eq:tb-hopxy}
\end{eqnarray}
\endnumparts (and analogously for
$\uuline{{t}}(\pm\hat{\bi{y}}\pm\hat{\bi{z}})$). The matrices related
to the $t^{2z}$ hopping parameter have the same form as
\eref{eq:tb-hopz} and \eref{eq:tb-hopx}. In the GFO distorted
structure, the matrix elements are also reduced by $(1-\eta_{t}^{s})$.

The Hund's rule coupling strength is determined using
\eref{eq:tb-hund} from the orbitally averaged spin splitting of the
Hamiltonian on-site diagonal matrix elements (for the $Pbnm$
structure)
\begin{equation}\label{eq:tb-J}
  J_{\mathrm{H}}=
  \case{1}{4}
  \left[
    \left(h_{11}^{0}+h_{22}^{0}\right)^{\downarrow}
    -\left(h_{11}^{0}+h_{22}^{0}\right)^{\uparrow}
  \right]\,.
\end{equation}
The (generally spin-dependent) JT coupling parameter $\lambda^{s}$ is
determined (from the JT($Q^{x}$) distorted structure) as
\begin{equation}\label{eq:tb-lambda}
  \lambda^{s}=
  \frac{\Delta\varepsilon^{s}}{2\vert{Q^x}\vert}\,,
\end{equation}
where the JT induced eigenvalue splitting $\Delta\varepsilon$ of the
$e_{g}$ subspace $2 \times 2$ on-site matrix is calculated as
\begin{equation}\label{eq:tb-deltaeps}
  \Delta\varepsilon=
  \left[{\left(h_{11}^{0}-h_{22}^{0}\right)^2+\left(2h_{12}^{0}\right)^2}\right]^{1/2}\,.
\end{equation}
The JT coupling $\lambda^{s}$ is effectively reduced due to the GFO
distortion mode (see~\cite{kovacik10} for more details) via the
$\eta_{\lambda}$ parameter calculated as
\begin{equation}\label{eq:tb-etalambda}
  \eta_{\lambda}=1-
  \frac{\Delta\varepsilon^{\uparrow}[Pbnm]}{\Delta\varepsilon^{\uparrow}[\mathrm{JT}(Q^x)]}
  \frac{\vert{\bi{Q}[\mathrm{JT}(Q^x)]}\vert}{\vert{\bi{Q}[Pbnm]}\vert}
  \,,
\end{equation}
where $\vert\bi{Q}\vert=\sqrt{(Q^x)^2+(Q^z)^2}$.

\end{document}